RESEARCH

# Social Engineering in Cybersecurity: A Domain Ontology and Knowledge Graph Application Examples

Zuoguang Wang[1,2*], Hongsong Zhu[1,2*], Peipei Liu[1,2] and Limin Sun[1,2]

**Abstract**

Social engineering has posed a serious threat to cyberspace security. To protect against social engineering attacks, a fundamental work is to know what constitutes social engineering. This paper first develops a domain ontology of social engineering in cybersecurity and conducts ontology evaluation by its knowledge graph application. The domain ontology defines 11 concepts of core entities that significantly constitute or affect social engineering domain, together with 22 kinds of relations describing how these entities related to each other. It provides a formal and explicit knowledge schema to understand, analyze, reuse and share domain knowledge of social engineering. Furthermore, this paper builds a knowledge graph based on 15 social engineering attack incidents and scenarios. 7 knowledge graph application examples (in 6 analysis patterns) demonstrate that the ontology together with knowledge graph is useful to 1) understand and analyze social engineering attack scenario and incident, 2) find the top ranked social engineering threat elements (e.g. the most exploited human vulnerabilities and most used attack mediums), 3) find potential social engineering threats to victims, 4) find potential targets for social engineering attackers, 5) find potential attack paths from specific attacker to specific target, and 6) analyze the same origin attacks.

## 1 Introduction

In the context of cybersecurity, social engineering describes a type of attack in which the attacker exploit human vulnerabilities (by means such as influence, persuasion, deception, manipulation and inducing) to breach the security goals (such as confidentiality, integrity, availability, controllability and auditability) of cyberspace elements (such as infrastructure, data, resource, user and operation). Succinctly, social engineering is a type of attack wherein the attacker exploit human vulnerability through social interaction to breach cyberspace security [1]. Many distinctive features make social engineering to be a quite popular attack in hacker community and a serious, universal and persistent threat to cyber security. 1) Compared to classical attacks such as password cracking by brute-force and software vulnerabilities exploit, social engineering exploits human vulnerabilities to bypass or break through security barriers, without having to combat with firewall or antivirus software by deep coding. 2) For some attack scenarios, social engineering can be as simple as making a phone call and impersonating an insider to elicit the classified information. 3) Especially in past decades when defense mainly focus on the digital domain yet overlooks human factors in security. As the development of security technology, classical attacks become harder and more and more attackers turn to social engineering. 4) Human vulnerabilities seem inevitable, after all, there is not a cyber system doesn't rely on humans or involve human factors on earth and these human factors are vulnerable obviously or can be largely turned into security vulnerabilities by skilled attackers. Moreover, social engineering threat is increasingly serious along with its evolution in new technical and cyber environment. Social engineering gets not only large amounts of sensitive information about people, network and devices but also more attack channels with the wide applications of So-

*Correspondence: wangzuoguang16@mails.ucas.ac.cn; zhuhongsong@iie.ac.cn
[1]School of Cyber Security, University of Chinese Academy of Sciences, Beijing, CN
[2]Beijing Key Laboratory of IoT Information Security Technology, Institute of Information Engineering, Chinese Academy of Sciences, Beijing, CN
Full list of author information is available at the end of the article cial Networking Sites (SNSs), Internet of Things (IoT), Industrial Internet, mobile communication and wearable devices. And large part of above information is open source, which simplifies the information gathering for social engineering. Social engineering becomes more efficient and automated by technology such as machine learning and artificial intelligence. As a result, a large group of targets can be reached and specific victims can be carefully selected to craft more creditable attack. The spread of social engineering tools decrease the threat threshold. Loose office policy (bring your own device, remote office, etc.) leads to the weakening of area-isolation of different security levels and creates more attack opportunities. Targeted, large-scale, robotic, automated and advanced social engineering attack is becoming possible [1].

To protect against social engineering, the fundamental work is to know what social engineering is, what entities significantly constitute or affect social engineering and how these entities relate to each other. Study [1] proposed a definition of social engineering in cybersecurity based on systematically conceptual evolution analysis. Yet only the definition is not enough to get insight into all the issue above, and further, to server as a tool for analyzing social engineering attack scenarios or incidents and providing a formal, explicit, reusable knowledge schema of social engineering domain.

Ontology is a term comes from philosophy to describe the existence of beings in the world and adopted in informatics, semantic web, knowledge engineering and Artificial Intelligence (AI) fields, in which an ontology is a formal, explicit description of knowledge as a set of concepts within a domain and the relationships among them (i.e. what entities exist in a domain and how they related). It defines a common vocabulary for researchers who need to share information and includes definitions of basic concepts in the domain and their relations [2]. In an ontology, semantic information and components such as concept, object, relation, attribute, constraints and axiom are encoded or formally specified, by which an ontology is machine-readable and has capacity for reasoning. In this way, ontology not only introduce a formal, explicit, shareable and reusable knowledge representation but also can add new knowledge about the domain.

Thus, we propose a domain ontology of social engineering to understand, analyze, reuse and share domain knowledge of social engineering.

**Organization:** Section 2 describes the the background material and methodology to develop domain ontology. Section 3 presents the material and ontology implementation. Section 4 is the result: domain ontology of social engineering in cybersecurity. Section 5 is the evaluation and application of the ontology and knowledge graph. Section 6 is the discussion. Section 7 concludes the paper.

## 2 Methodology to develop domain ontology

There is no single correct way or methodology for developing ontologies [2]. Since ontology design is a creative process and many factors will affect the design choices, such as the potential applications of the ontology, the designer's understanding and view of the domain, different domain features, anticipations of the ontology to be more intuitive, general, detailed, extensible and / or maintainable.

In this paper, we design the methodology to develop domain ontology of social engineering based on the method reported in work [2] with some modification. Protégé 5.5.0 [3] is used to edit and implement the ontology. It should be noted that "entity" in real word are described as "concept" in ontology and "class" in Protégé; "relation" is described as "object property" in Protégé. The methodology is described as Figure 1.

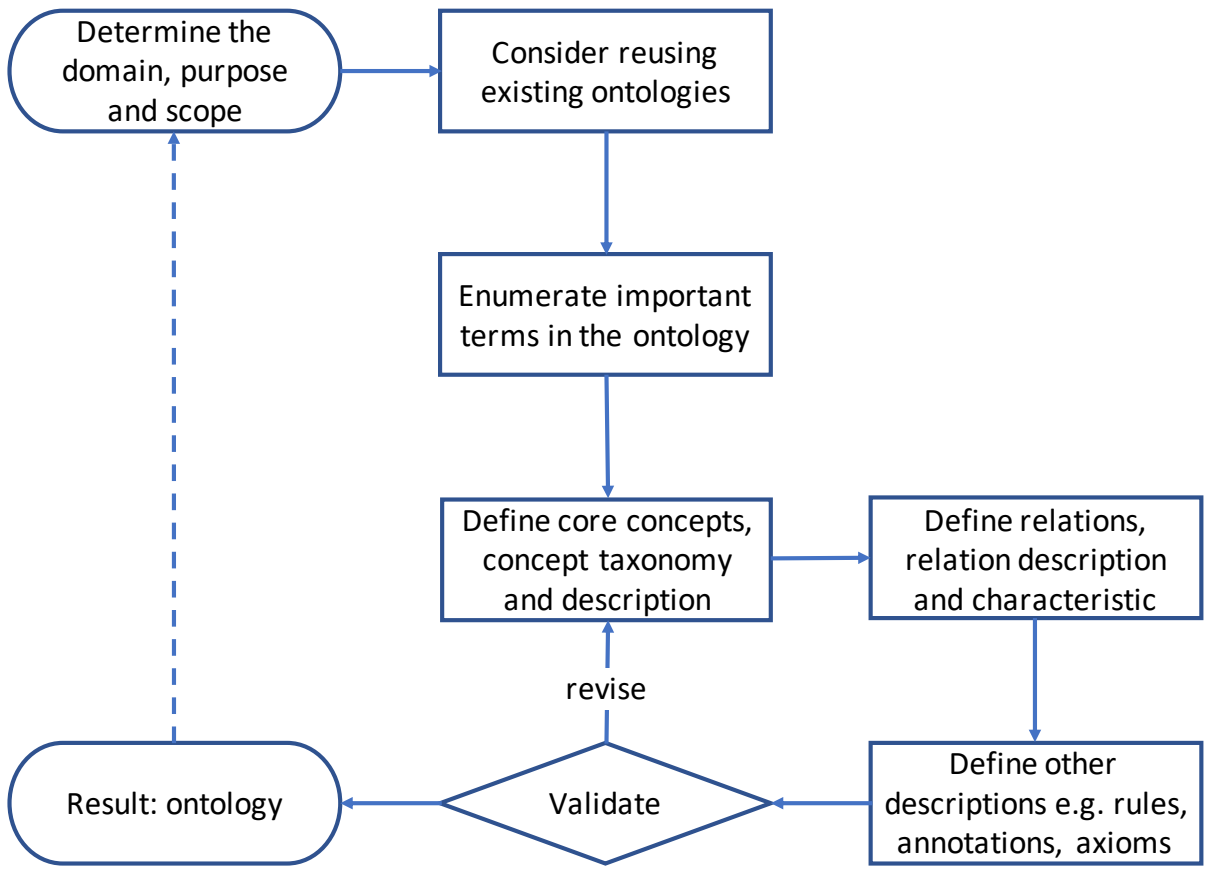

**Figure 1** Overview of methodology to develop domain ontology of social engineering

(1) Determine the domain, purpose and scope.

As described before, the domain of the ontology is social engineering in cybersecurity. The purpose of the ontology, i) for design is to present what entities significantly constitute or affect social engineering and how these entities relate to each other, ii) and for application is to server as a tool for understanding social engineering, analyzing social engineering attack scenarios or incidents and providing a formal, explicit, reusable knowledge schema of social engineering domain. Thus, social engineering itself as a type of attack, measures regarding social engineering defense will not be included here although they are important. Defense will be the theme in our future work.

(2) Consider reusing existing ontologies.

We did a systematic literature survey on social engineering and accumulated a literature database which contains 450+ studies from 1984.9 (time of the earliest literature available where the term "social engineering" was found in cybersecurity [1]) to 2020.5.[1] Few work focus on the social engineering ontology, yet a lot of terms can be obtained from literature survey.

(3) Enumerate important terms in the ontology.

"Initially, it is important to get a comprehensive list of terms without worrying about overlap between concepts they represent, relations among the terms ..." [2]. These terms are useful to intuitively and quickly get a sketchy understanding on a domain, and helpful to develop a core concepts set after due consideration. A total of 350 relevant terms are enumerated from the literature database mentioned in (2). Table 1 shows these terms in a compact layout by length order. [2]

The next two steps are the most important steps in the ontology design process [2].

(4) Define core concepts, concept taxonomy and description.

In work [2], this step is to create the class hierarchy for a single concept "Wine". However, the "class, sub-class" hierarchy is a structure typically used to classification, in which only the relation "is a" or "is type of" is described. This is not the purpose of this paper. Thus, differently, we define a set of concepts for entities which significantly constitute or affect social engineering domain and discuss their taxonomy. Then, we define more expressive relations among concepts in next step.

For each core concept, a definition is provided and relevant synonym terms are mentioned, to facilitate the reuse and sharing of domain knowledge. For example, attacker (a.k.a. social engineer) is the party to conduct social engineering attack; it can be an individual or an organization, and internal or external. In Protégé, these concepts are edited in the "Classes" tab. Two Classes "Attacker" and "Social Engineer" are created and because they represent the same class (concept), a description (class axiom) "Equivalent To" is set between them in the "Description" tab. As Figure 2 shows.

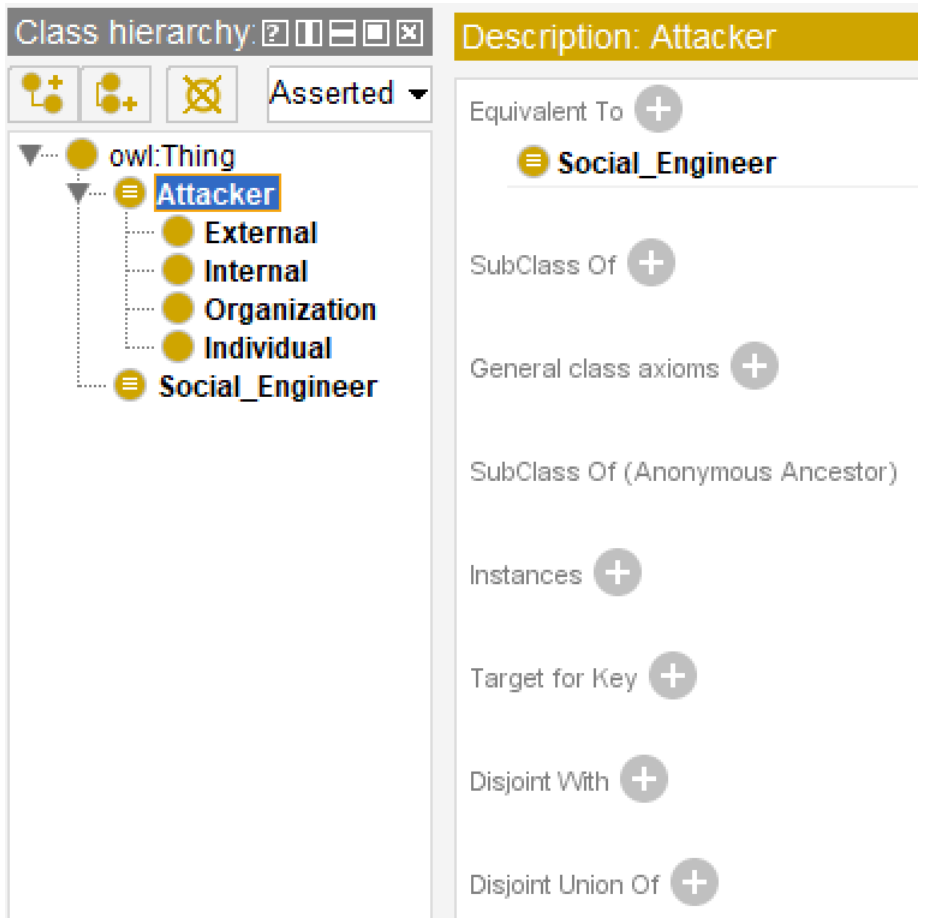

**Figure 2** Edit concepts and their description

(5) Define relations, relation description and characteristic.

This step we create the relations among concepts based on their definitions. Some relations directly expressed in the definition while some may be implicit and need a explicit description. For example, attack motivation is the factors that motivate (incent, drive, cause or prompt) the attacker to conduct a social engineering attack; thus, a concise relation "motivate" from "attack motivation" to "attacker" can be created. And to be more compatible, two sub-relation "incent" and "drive" or another equivalent relation can be added. In Protégé, these relations are edited in the "Object properties" tab. For above example, "motivate" as an Object property is created; "Attack Motivation" is its Domain and "Attacker" is its Range. Because it represents that a class points to another different class, the relation characteristic "Irreflexive" is set. As Figure 3 shows.

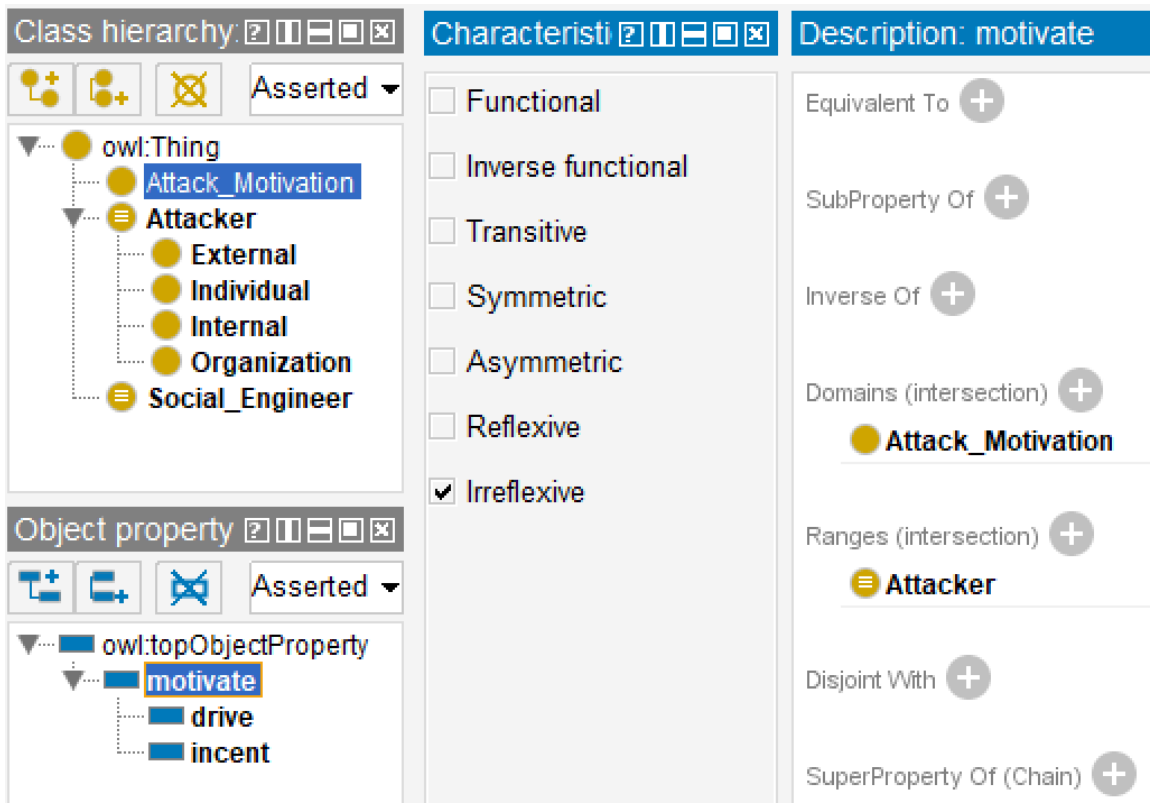

**Figure 3** Edit relations, relation description and characteristic

(6) Define other descriptions.

Besides above, other descriptions can be added, such as annotations, axioms, rules. Examples are as follows. For class "Attacker", its definition can be added as a comment in Annotations tab with "rdfs:comment", to facilitate conceptual understanding and later debug. Axioms are statements that are asserted to be true. For

[1]The literature database was submitted as supplementary material for review.
[2]Term lists organized by alphabetical order and semantic groups were submitted as supplementary material for review.

**Table 1** Terms related to social engineering in cybersecurity

| | | | | | | | | | |
|---|---|---|---|---|---|---|---|---|---|
| APT | groups | Facebook | Instagram | person name | attack vector | company partner | reciprocity norm | mobile application | posts in social media |
| fun | hacker | flattery | integrity | server name | central route | confidentiality | religion belief | network disruption | similarity and liking |
| war | hubris | gluttony | interests | take effect | desk sniffing | controllability | shoulder surfing | penetration tester | vulnerability exploit |
| XSS | induce | humility | prejudice | attach files | eavesdropping | decision making | social relations | perform attack | intellectual challenge |
| bias | letter | identity | principle | attack model | employee name | deindividuation | website phishing | source credibility | internal phone numbers |
| card | medium | kindness | secretary | attack skill | impersonation | direct approach | adjacent overhear | telephone operator | malicious popup window |
| CSRF | motive | laziness | self-love | auditability | item dropping | dumpster diving | attack motivation | trust relationship | obtain physical access |
| envy | piston | LinkedIn | terrorism | availability | launch attack | economic profit | behavioral habits | authoritative voice | reputation destruction |
| fear | spying | openness | user name | carelessness | name-dropping | fake mobile app | computer operator | cultural disruption | social engineering bot |
| goal | target | password | compliment | email footer | reverse sting | group influence | conscientiousness | data exfiltration | social exchange theory |
| hoax | victim | phishing | conformity | email format | security risk | instant message | data modification | denial of service | software vulnerability |
| KeeK | baiting | phreaker | contractor | equivocation | self interest | office snooping | deceptive website | emotion and feeling | thought and expression |
| lust | charity | pleasure | diffidence | extraversion | time pressure | craft attack | drive-by download | executive assistant | vulnerability analysis |
| name | clients | politics | excitement | face-to-face | trojan attack | self-disclosure | drive-by-pharming | individual attacker | portable storage drives |
| scam | disgust | RFID tag | flirtation | friendliness | trojan device | social disorder | external attacker | intuitive judgement | questionnaire surveying |
| SNSs | friends | scarcity | heuristics | inexperience | vulnerability | social engineer | external pressure | neurophysiological | Social Networking Sites |
| anger | Google+ | smishing | job title | ingratiation | watering hole | social software | facial expression | organizational logo | the quest for knowledge |
| cloud | hobbies | software | moral duty | interruption | attack pattern | thoughtlessness | instant messenger | unauthorized access | administrative assistant |
| dread | manager | strategy | motivation | intimidation | attack purpose | application name | internal attacker | cognitive dissonance | organizational structure |
| email | manuals | surprise | persuasion | IP addresses | build relation | attack framework | IT infrastructure | disgruntled employee | voice mail systems vendor |
| greed | partner | sympathy | pretending | manipulation | financial gain | attack technique | network intrusion | facial action coding | commitment and consistency |
| guilt | phisher | trailing | pretexting | masquerading | framing effect | bystander effect | personal interest | interesting malwares | human resources department |
| lingo | picture | trashing | road apple | new employee | identity thief | confidence trick | physical presence | network interception | reverse social engineering |
| photo | profile | weakness | attack goal | piggybacking | image spoiling | data destruction | physical sabotage | rapport relationship | social responsibility norm |
| prank | purpose | authority | attack path | quid pro quo | mobile devices | data fabrication | political purpose | system administrator | diffusion of responsibility |
| Skype | QR code | bluetooth | attack plan | receptionist | mobile website | e-mail addresses | social validation | Voice over IP (VoIP) | short message service (SMS) |
| sloth | revenge | calendars | connections | social proof | movable device | effect mechanism | sports fanaticism | accounting department | Elaboration Likelihood Model |
| trick | sadness | credulity | distraction | stereotyping | pop-up windows | financial return | technical support | attacker organization | computer hardware manufacturer |
| video | tension | curiosity | elicitation | thinking set | security guard | foot-in-the-door | attack consequence | competitive advantage | computer software manufacturer |
| weibo | Twitter | deception | gullibility | trust theory | social network | IT professionals | creating confusion | fixed-action patterns | telephone system administrator |
| wrath | vishing | happiness | helpfulness | agreeableness | spear phishing | mental shortcuts | employee functions | impression management | low level of need for cognition |
| apathy | website | help desk | indifferent | attack medium | urgent request | micro expression | fake business card | information gathering | increasing the number of friends |
| attack | whaling | ignorance | information | attack method | attack approach | peripheral route | family information | instant communication | IVR (Interactive Voice Response) |
| awards | attacker | impulsion | neuroticism | attack target | attack strategy | person-to-person | gather information | language and thinking | interpersonal deception theory (IDT) |
| Flickr | courtesy | influence | overloading | attack threat | attacker group | phone, telephone | habitual behaviors | organizational policy | integrative model of organizational trust |

relation "motivate", we can create an inverse relation "motivated by" and then set the description (object property axiom) "Inverse Of" against "motivate", to facilitate the knowledge retrieval like "attacker is motivated by certain attack motivation". Ontology can also generate new knowledge by reasoning with rules. Assume that "different attackers are regarded as from the same attack organization if they motivated by the same motivation and attack the same victim", then the following rule can be defined to implement the reasoning. *Rule:* $motivate(?m, ?a) \wedge attack(?a, ?v) \wedge motivate(?m, ?b) \wedge attack(?b, ?v) \wedge differentFrom(?a, ?b) \rightarrow same_attack_organization(?a, ?b)$. As Figure 4 shows.

(7) Validate and revise.

After defining the concepts, relations and related descriptions, a domain ontology is created. Yet it is initial and imperfect. Minor mistakes such as misplacement and typing error may be occurred when large amount of items existed. Illogical or contradictory descriptions may be defined. Some class, relations or descriptions may be absent or superfluous. Thus, an iterative process is necessary for ontology development, validation and revision.

By virtue of the ontology is formal and explicit encoded, any faults that cause logical inconsistency can be found. The built-in reasoner HermiT is used for this reasoning validation. Further, we create instances as the actual data to conduct a deductive validation, as Figure 4 shows. This is an intuitive method to test whether the ontology (e.g. the rules) is effective, and it also provides a way helpful to adjust descriptions and revise the ontology to achieve the purpose previously.

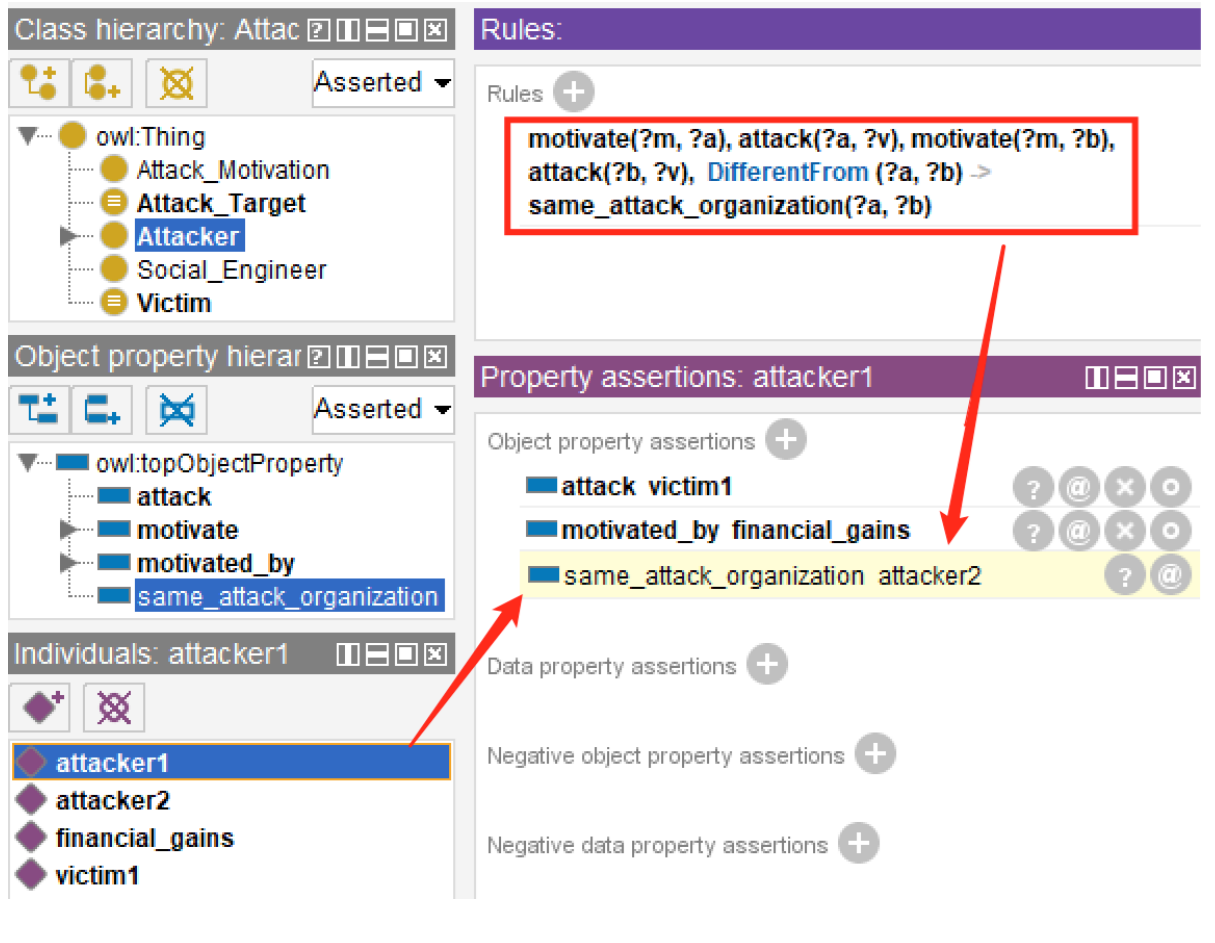

**Figure 4** Define and apply rules to knowledge reasoning

(8) Result: Ontology.

Finally, a domain ontology of social engineering is developed after iterative revision and validation.

## 3 Material and ontology implementation

The background material regarding literature and terms have been mentioned in Section 2 and we will not repeat them here. This section presents the key material and procedures for the ontology implementation, i.e. defining the concepts, relations and other descriptions related.

### 3.1 Define core concepts in the domain ontology

This subsection details 11 core concepts corresponding to entities that significantly constitute or affect social engineering domain. For each concept, the concept definition, synonym term, taxonomy and some other properties are described. Figure 5 shows these entities (concepts). The circular arrow represents an approximate attack cycle for typical attack scenarios: 1) the attacker motivated by certain factors 2) to gather specific information, formulate attack strategy, craft attack method 3) and then through certain medium the attack method is performed and the attack target is interacted with 4) to exploit their vulnerabilities which take effect and lead to attack consequences; 5) the consequence feed back to the attack goal predetermined to satisfy the attack motivation.

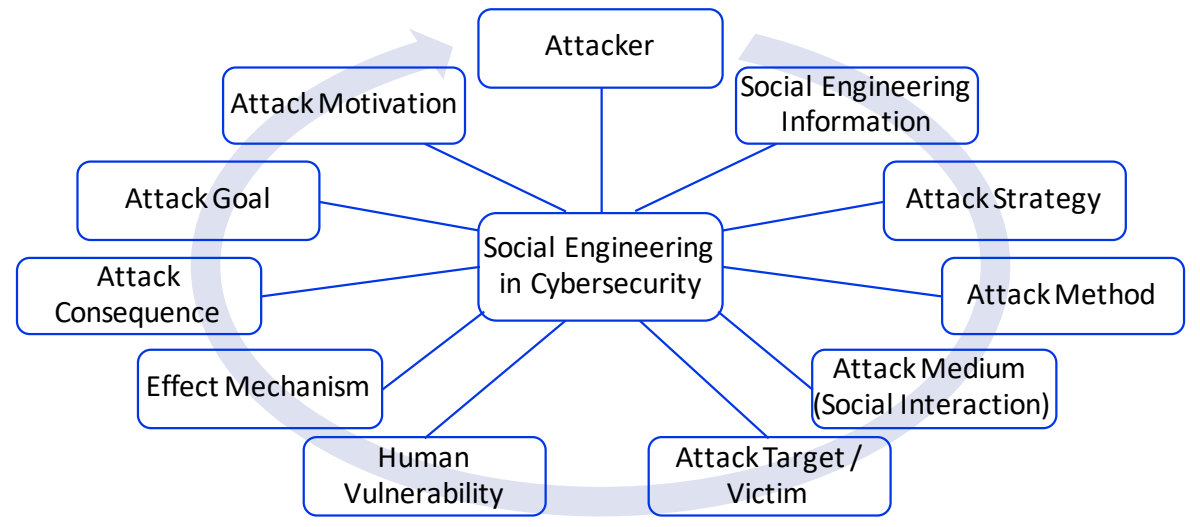

**Figure 5** Core entities (concepts) in social engineering domain

#### *3.1.1 Attacker*

For social engineering, the attacker (a.k.a. social engineer) is the party to conduct a social engineering attack; it is typically motivated by certain factors discussed in Section 3.1.2. Social engineering attackers appear in various forms in reality, such as hackers, phreakers, phishers, disgruntled employees, identity thieves, penetration testers, script kiddies, malicious users. Different criteria can also be used for the attacker's taxonomy. The attacker identified as an individual person is familiar to the public, yet it does not have to be an individual. The attacker can also be a group or an organization. The attacker can be a real person, or a virtual human role (e.g. a bot), and it can be from internal or external. Figure 6.

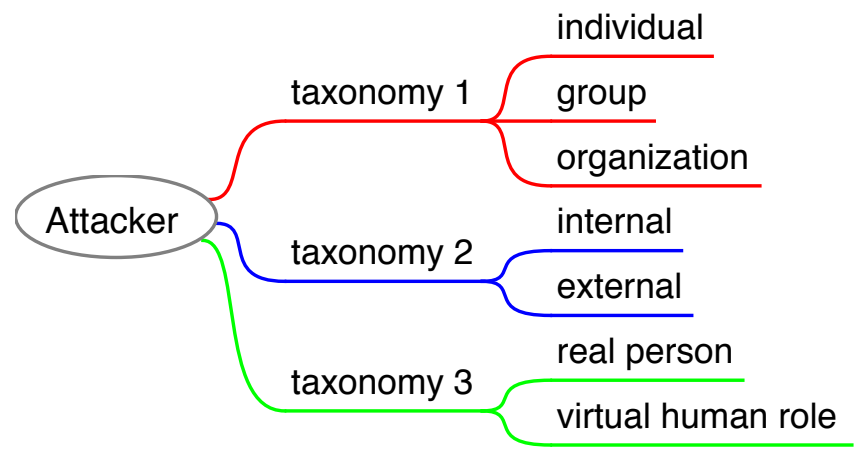

**Figure 6** Taxonomy of attacker (social engineer)

#### *3.1.2 Attack Motivation*

Attack motivation is the factors that motivate (incent, drive, cause or prompt) the attacker to conduct a social engineering attack. It can be intrinsic or extrinsic. Considering that this simple taxonomy does not seem to be significantly helpful to the social engineering analysis, a common list of attack motivations in social engineering may be more intuitive. It includes but is not limited to: 1) financial gain [4], 2) competitive advantage [5], 3) revenge [4], 4) external pressure, 5) personal interest, 6) intellectual challenge, 7) increasing followers or friends in SNSs, 8) image spoiling (denigration, reputation destruction, stigmatization), 9) prank, 10) fun or pleasure, 11) politics, 12) war, 13) religious belief, 14) fanaticism, 15) social disorder, 16) cultural disruption [6], 17) terrorism, 18) espionage, 19) security test.

#### *3.1.3 Attack Goal and Object*

The attack goal (a.k.a. attack purpose) is something that the attacker wants to achieve by specific attack methods so that the attack motivation can be satisfied. For social engineering, it is some kinds of breaching against cyberspace security. In general, to breach cyberspace security is to breach the security goals (confidentiality, integrity, availability, controllability, auditability, etc.) of the four basic elements of cyberspace (i.e. attack object) [1]. These four basic elements are Carrier (the infrastructure, hardware and software facilities of cyberspace), Resources (the objects, data content that flows through the cyberspace), Subjects (the main body roles and users, including human users, organizations, equipment, software, websites, etc.), and Operations (all kinds of activities of processing Resources, including creation, storage, change, use, transmission, display, etc.) [7, 8]. For complex attack scenarios, there may be sub-goals (precondition) exist, which themselves may not breach the cybersecurity.

Social engineering attack goal includes but is not limited to: 1) network intrusion, interception or disruption, 2) gain unauthorized access to information or systems, 3) denial of service, 4) data exfiltration, modification, fabrication or destruction, 5) infrastructure sabotage, 6) obtain physical access to restricted areas. Thus, it can be simply classified as above categories or use other taxonomies as Figure 7 shows.

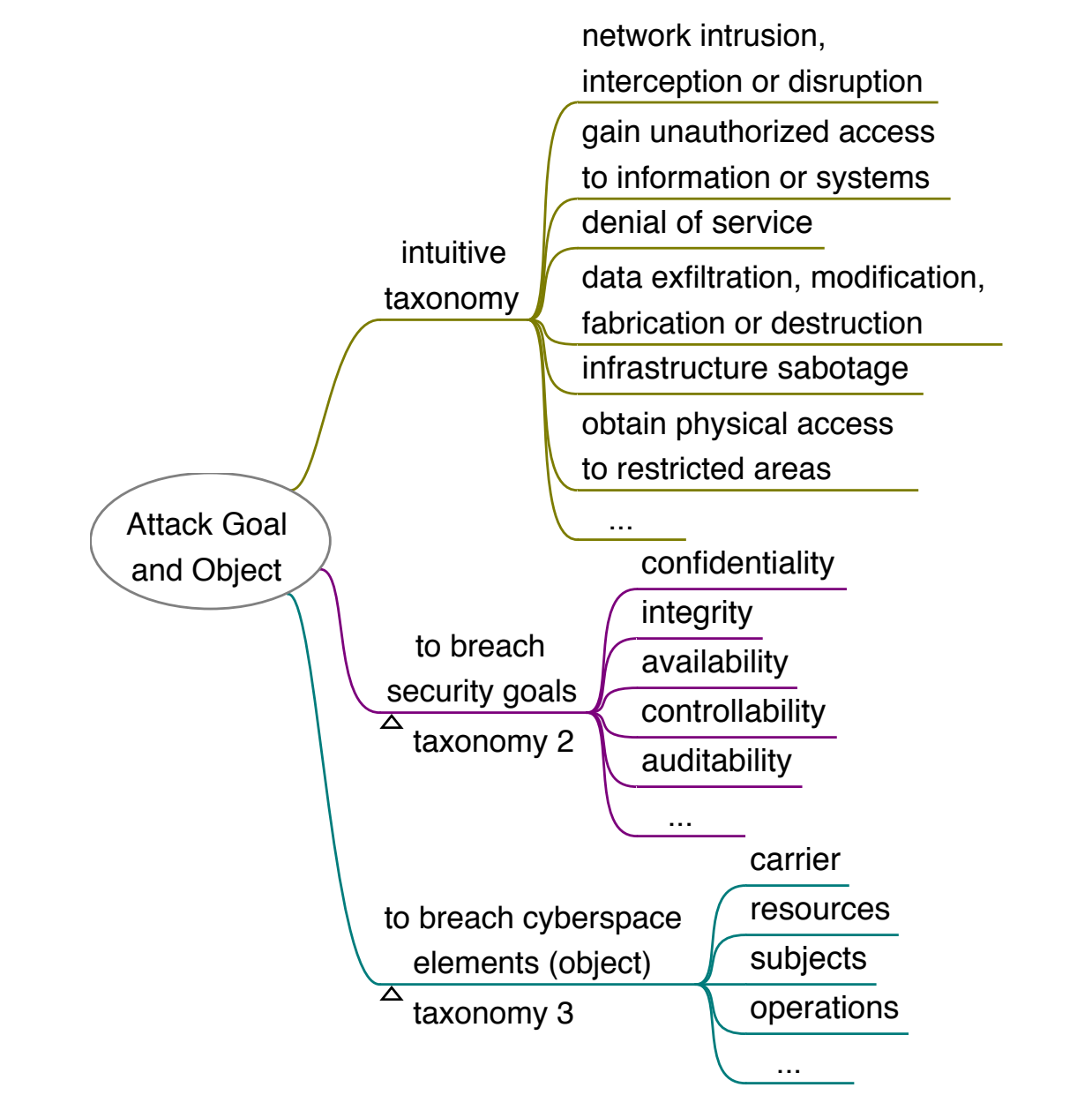

**Figure 7** Taxonomy of social engineering attack goal

### *3.1.4 Social Engineering Information*

In many attack scenarios, the success of social engineering relies heavily on the information gathered, such as personal information of the targets (victims), organization information, network information, social relation information. In a broad sense, every bit of information posted publicly or leaked in cyberspace or in reality might provide attackers the resource, such as to learn the environment, to discover targets, to find vulnerable human factors and cyber vulnerabilities, to formulate attack strategy, and to craft attack methods. This is also a feature of social engineering compared with classical computer attack. Thus, this paper use "social engineering information" to represent any information that helps the attacker to conduct a social engineering attack.

Social engineering information includes but is not limited to: 1) person name, 2) identity 3) photograph, 4) habits and characteristics, 5) hobbies or interests, 6) job title, 7) job responsibility, 8) schedule, 9) routines, 10) new employee, 11) organizational structure, 12) organizational policy, 13) organizational logo, 14) company partner, 15) lingo, 16) manuals, 17) interpersonal relations, 18) family information, 19) profile in SNSs, 20) posts in social media, 21) connections in SNSs, 22) SNSs group information, 23) (internal) phone numbers, 24) email information (address, format, footer, etc.), 25) username, 26) password, 27) network information, 28) computer name, 29) IP addresses, 30) server name, 31) application information, 32) version information, 33) hardware information, 34) IT infrastructure information, 35) building structure, 36) location information.

Figure 8 presents a taxonomy based on what space the information describes, in which the last level may be more intuitive. Other taxonomies can be also workable, such as publicly accessible information, restricted information; personal information, social relations information and other various environments (cyber, cultural, physical) information.

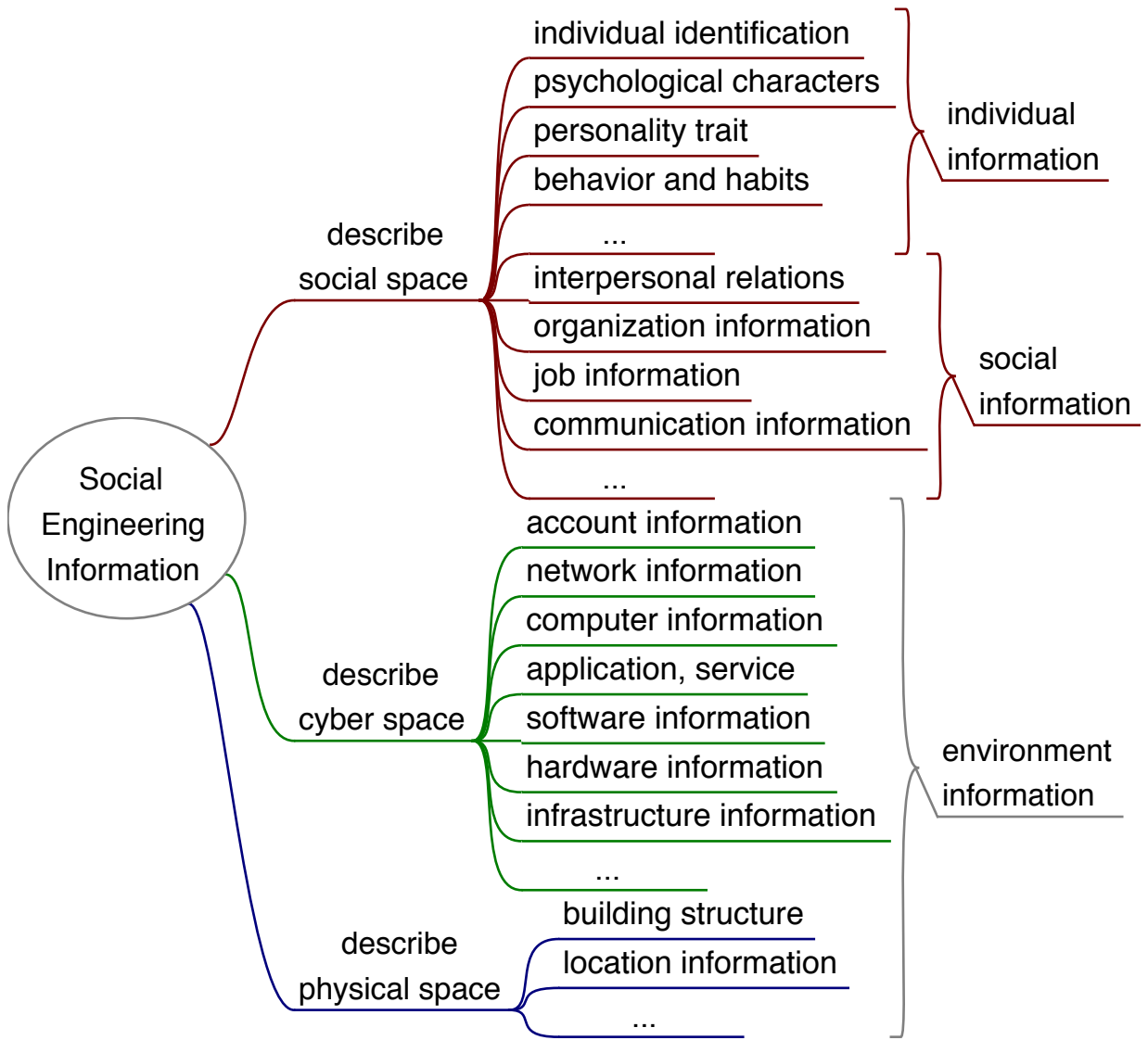

**Figure 8** Taxonomy of social engineering information

### *3.1.5 Attack Strategy*

Attack strategy is a plan, pattern, or guidance of actions formulated by the attacker for certain attack goal. It is necessary especially for complex social engineering attacks. Usually, social engineering attackers formulate the attack strategy based on their comprehensive understanding on the attack situation, such as resources, environments, targets, vulnerabilities and mediums. There are two common social engineering strategies in literature: forward (usual) strategy and reverse strategy. In forward attack strategy, the attacker directly contacts the targets and delivers attack payloads to them, waiting the targets to trigger the attack and be compromised. However, in reverse social engineering, the targets are prompted to contact the attacker actively for a request or help, and the attacker usually pretends to be a party of legitimate, authoritative, expert or trustworthy in advance. As a result, a higher degree of trust is established and the targets are more likely to be attacked. E.g. The attacker first makes a network failure and then pretends to be a technical support staff; when the targets seek for a help, the attacker convinces them with certain excuses into revealing the password or installing a malicious software.

From the duration perspective, attack strategy can be persistent strategy or short-term strategy. Some other categories are also helpful to label the attack strategies, as Figure 9 shows.

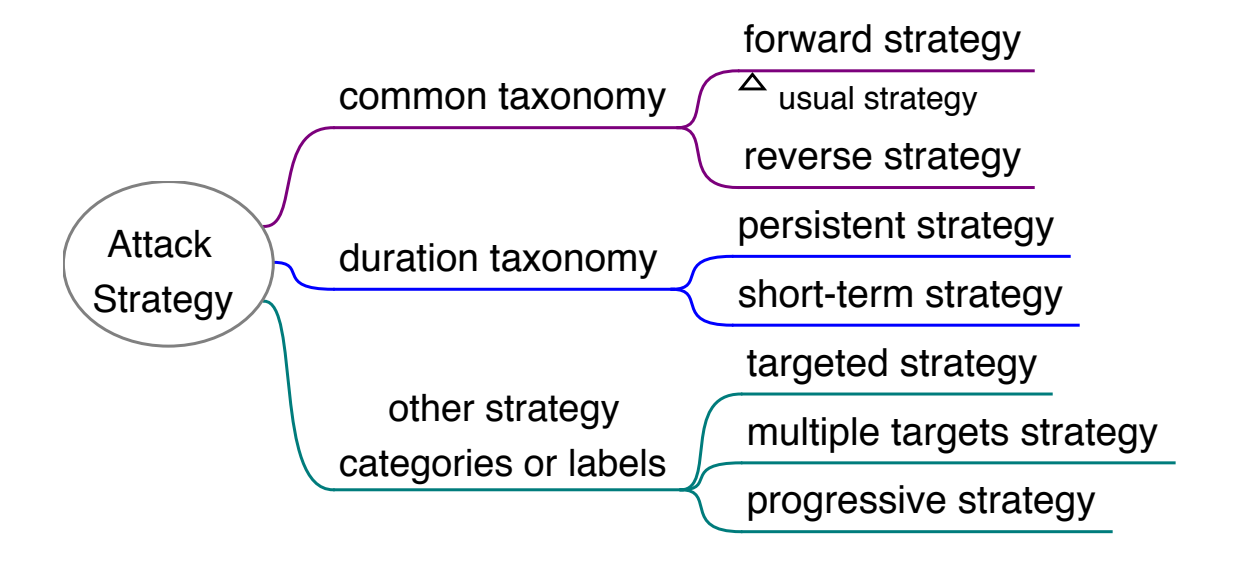

**Figure 9** Taxonomy of social engineering attack strategy

### *3.1.6 Attack Method*

When the attack strategy existed, attack method is generally according to or guided by it. Attack method is the way, manner or means of carrying an attack out; the attacker crafts and performs it to achieve specific attack goal. Synonyms such as attack vector, attack technique and attack approach are used to convey the same meaning. A common taxonomy in literature is to divide social engineering attacks into human-based and computer-based (or technology-based) [9–13]. Figure 10 (right) presents 20 attack method instances, in which some methods such as influence, deception, persuasion, manipulation and induction also describe skills frequently used in other methods. In many attack scenarios, multiple social engineering methods can be jointly used; classical attack methods that exploit non-human-vulnerabilities might also be combined to perform social engineering attacks. Besides, there are many auxiliary tricks or cunning actions may be utilized in different methods to assist the attack (e.g. to obtain trust, influence or deceive the targets). Figure 10 shows the overview of these categories and the corresponding instances. It is a non-exhaustive list and it seems impossible to enumerate all the social engineering attack methods, since new attack methods are emerging as the development of cyber technology, the evolution of environment and attackers' creation.

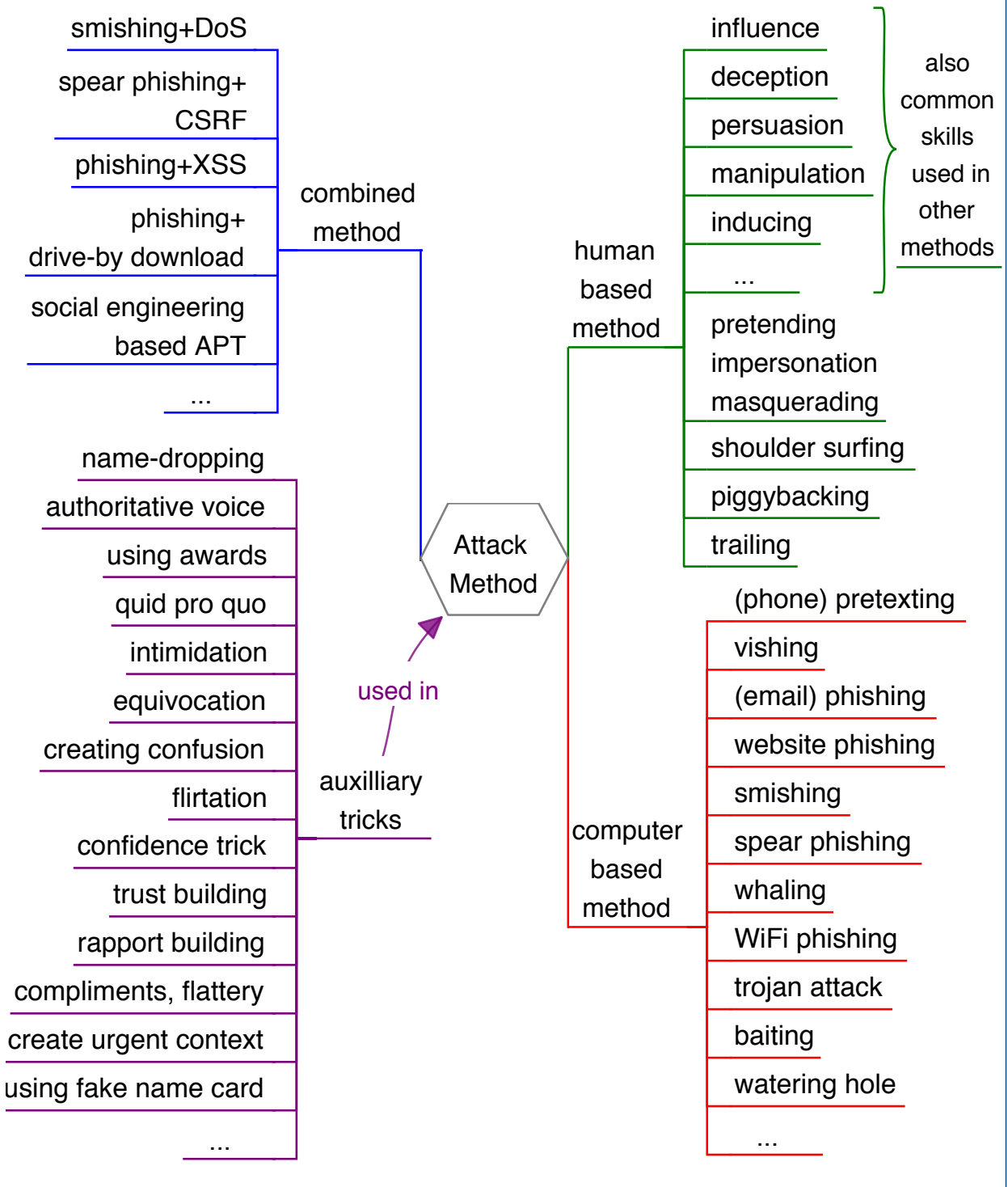

**Figure 10** Taxonomy of social engineering attack method

### *3.1.7 Attack Target, Victim*

Attack target is the party to suffer a social engineering attack and bring about an attack consequence. The attacker applies attack method to the targets, and they become victims once their vulnerabilities were exploited. For attackers, anyone helpful to achieve the attack goal is a potential attack target. And the attacker might select multiple targets in some attack scenarios. The potential attack targets include but is not limited to: 1) new employees, 2) secretaries, 3) help desk, 4) technical support, 5) system administrators, 6) telephone operators, 7) security guards, 8) receptionists, 9) contractors, 10) clients, 11) partners, 12) managers, 13) executive assistants, 14) manufacturers, 15) vendors [14]. Similar to the attacker, attack target can be an individual, a group or an organization; a real person or a virtual human role; from internal or external. As Figure 11 shows.

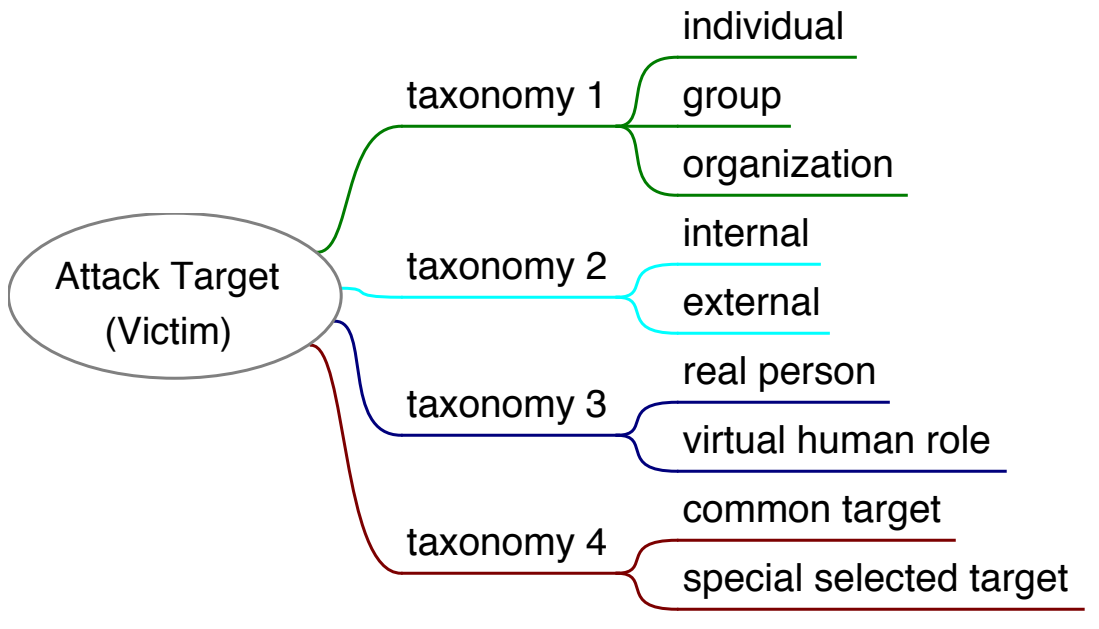

**Figure 11** Taxonomy of social engineering attack target

### *3.1.8 Social Interaction and Attack Medium*

Social engineering is a type of attack involves social interaction which is defined as the communication between or joint activity involving two or more human roles [1]. It covers the interpersonal interaction in the real world and user interaction in cyberspace. Attack medium is not only the entity so that the social interaction can implement (through which the target is contacted), but also the substance or channel through which attack methods are carried out. In some social engineering attacks, several different mediums might be used. E.g. The attacker deceives the target through phone to receive an important document, and then carry out phishing attack in the email.

The taxonomies of social interaction can be various according to different criteria. It can be direct (e.g. face to face in the real world) or indirect (e.g. email), real-time (e.g. phone talking) or non-real-time (e.g. email), active or passive (e.g. reverse social engineering). As Figure 12 shows.

The attack mediums include but is not limited to: 1) the real world, 2) attach files, 3) letter, 4) manual, 5) card, 6) picture, 7) video, 8) RFID tag, 9) QR code, 10) phone, 11) email, 12) website, 13) software, 14) Bluetooth, 15) pop-up window, 16) instant messenger, 17) cloud service, 18) Voice over IP (VoIP), 19) portable storage drives, 20) short message service (SMS), 21) mobile communication devices, 22) SNSs.

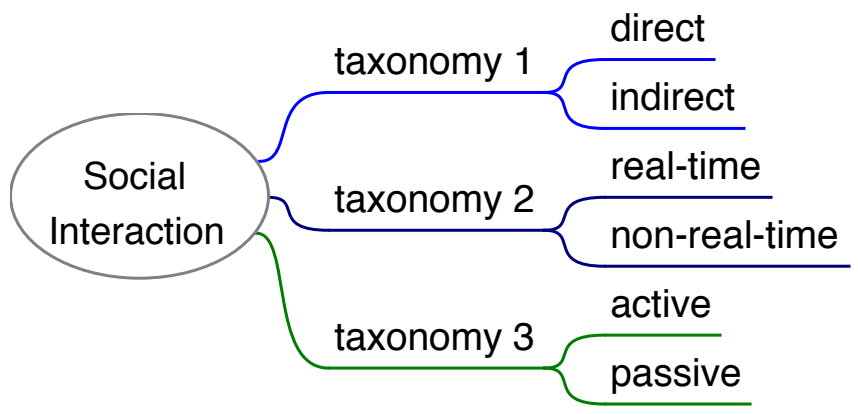

**Figure 12** Taxonomy of social interaction in social engineering

### *3.1.9 Human Vulnerability*

Human vulnerability is the human factor exploited by the attacker to conduct a social engineering attack through various kinds of attack methods. This is a distinctive attribute of social engineering compared to classical computer attacks. For social engineering, other types of vulnerability (e.g. software vulnerabilities) can be exploited together with human vulnerability, yet they are non-necessary [1]. A wide range of human factors can be exploited in social engineering, and a skilled social engineer (attacker) can transform common or inconspicuous human factors into security vulnerabilities exploitable in specific attack scenarios.

In general, human vulnerabilities in social engineering fall into four aspects: 1) cognition and knowledge, 2) behavior and habit, 3) emotion and feeling, and 4) psychological vulnerabilities. And the psychological vulnerabilities can be further divided into three levels: 1) human nature, 2) personality trait and 3) individual character from the evolution perspective of human wholeness to individuation [15]. Following is a non-exhaustive list of human vulnerabilities, which contains 43 instances of these six categories.

- Cognition and Knowledge (8 instances): ignorance, inexperience, thinking set and stereotyping, prejudice / bias, conformity, intuitive judgement, low level of need for cognition, heuristics and mental shortcuts.
- Behavior and Habit (4 instances): laziness / sloth, carelessness and thoughtlessness, fixed-action patterns, behavioral habits / habitual behaviors.
- Emotions and Feelings (11 instances): fear / dread, curiosity, anger / wrath, excitement, tension, happiness, sadness, disgust, surprise, guilt, impulsion, fluke mind.
- Human nature (6 instances): self-love, sympathy, helpfulness, greed, gluttony, lust.
- Personality traits (5 dimensions): conscientiousness, extraversion, agreeableness, openness, neuroticism.
- Individual characters (9 instances): credulity / gullibility, friendliness, kindness and charity, courtesy, humility, diffidence, apathy / indifferent, hubris, envy.

### *3.1.10 Effect Mechanism*

Social engineering effect mechanism describes the structural relation that what, why or how specific attack effect (consequence) corresponds to specific human vulnerability, in specific attack situation [15]. Given the attack scenarios and human vulnerabilities, it explains or predicts the attack consequence. E.g. Impression management theory and reciprocity norm explain why new employees (inexperience, helpfulness, etc.) are more vulnerable to give up their username and password to technical support staffs pretended by the attacker, who helps to resolve their network failure first and then request an information disclosure with certain excuses. Social engineering effect mechanisms involve lots of principles and theories in multiple disciplines such as sociology, psychology, social psychology, cognitive science, neuroscience and psycholinguistics. Study [15] summarizes six aspects of social engineering effect mechanisms: 1) persuasion, 2) influence, 3) cognition, attitude and behavior, 4) trust and deception, 5) language, thought and decision, 6) emotion and decision-making. Following is a non-exhaustive list of effect mechanisms, which contains 38 instances of these six aspects.

- Persuasion (7 instances): similarity & liking & helping in persuasion, distraction in persuasion and manipulation, source credibility and obey to authority, the central route to persuasion, the peripheral route to persuasion, Elaboration Likelihood Model of persuasion, recipient's need for cognition in persuasion.
- Influence (8 instances): group influence and conformity, normative influence (social validation), informational influence (social proof), social exchange theory, reciprocity norm, social responsibility norm, moral duty, self-disclosure and rapport relation building.
- Cognition, Attitude and Behavior (9 instances): impression management theory, cognitive dissonance, commitment and consistency, foot-in-the-door effect, diffusion of responsibility, bystander effect, deindividuation in group, time pressure and thought overloading, scarcity: perceived value and fear arousing.
- Trust and Deception (5 instances): trust and take risk, factor affecting trust, factor affecting deception, integrative model of organizational trust, interpersonal deception theory (IDT).
- Language, Thought and Decision (4 instances): relation between language and thinking, framing effect and cognitive bias, language invoke confusion: induce and manipulation, indirectness of thought and negative conception expression in language.
- Emotion and Decision-making (5 instances): neurophysiological mechanism of emotion & decision, emotion and feelings influence decision making, facial expression & deception leakage, facial action coding, micro expression identify and deception detecting.

### *3.1.11 Attack Consequence*

Attack consequence is something that follows as a result or effect of a social engineering attack. The attacker feed it back to the attack goal to decide whether a further attack is required. The taxonomy of attack consequence is similar with the taxonomy of attack goal, as Figure 13 shows.

Due to the subclass name in protégé will be converted to node labels in later knowledge graph, considering the intuitive demonstration and data feature, multiple different taxonomies can be used to assist knowledge analysis. Figure 14 (left) shows the implementation of concepts defined above. Table 2 shows the related concepts descriptions set as class axioms in protégé yet not reflected in the Figure 14. [3]

[3] The implementation file was submitted as supplementary material (SEiCS-Ontology+instances.owl) for review.

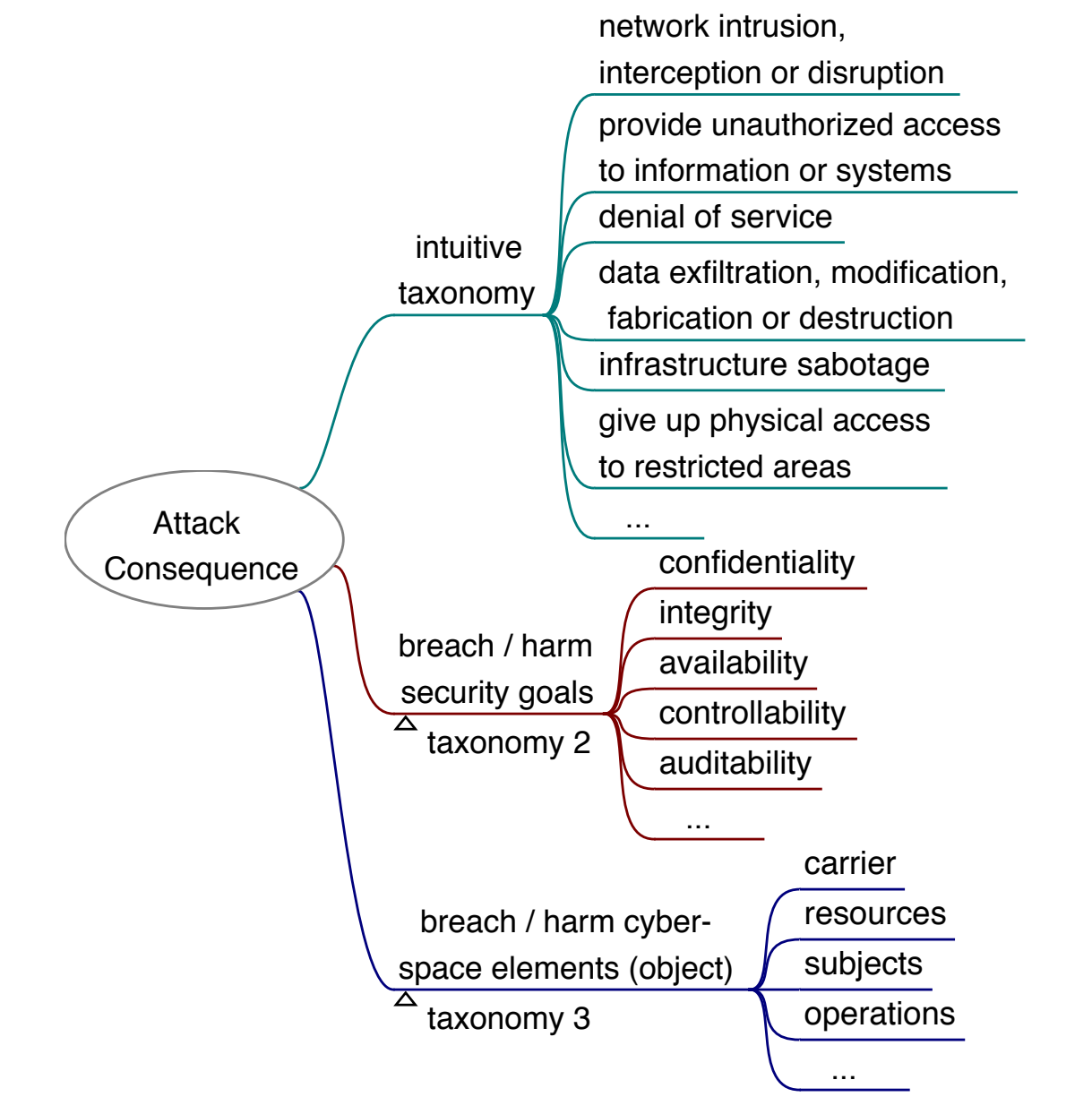

**Figure 13** Taxonomy of social engineering attack consequence

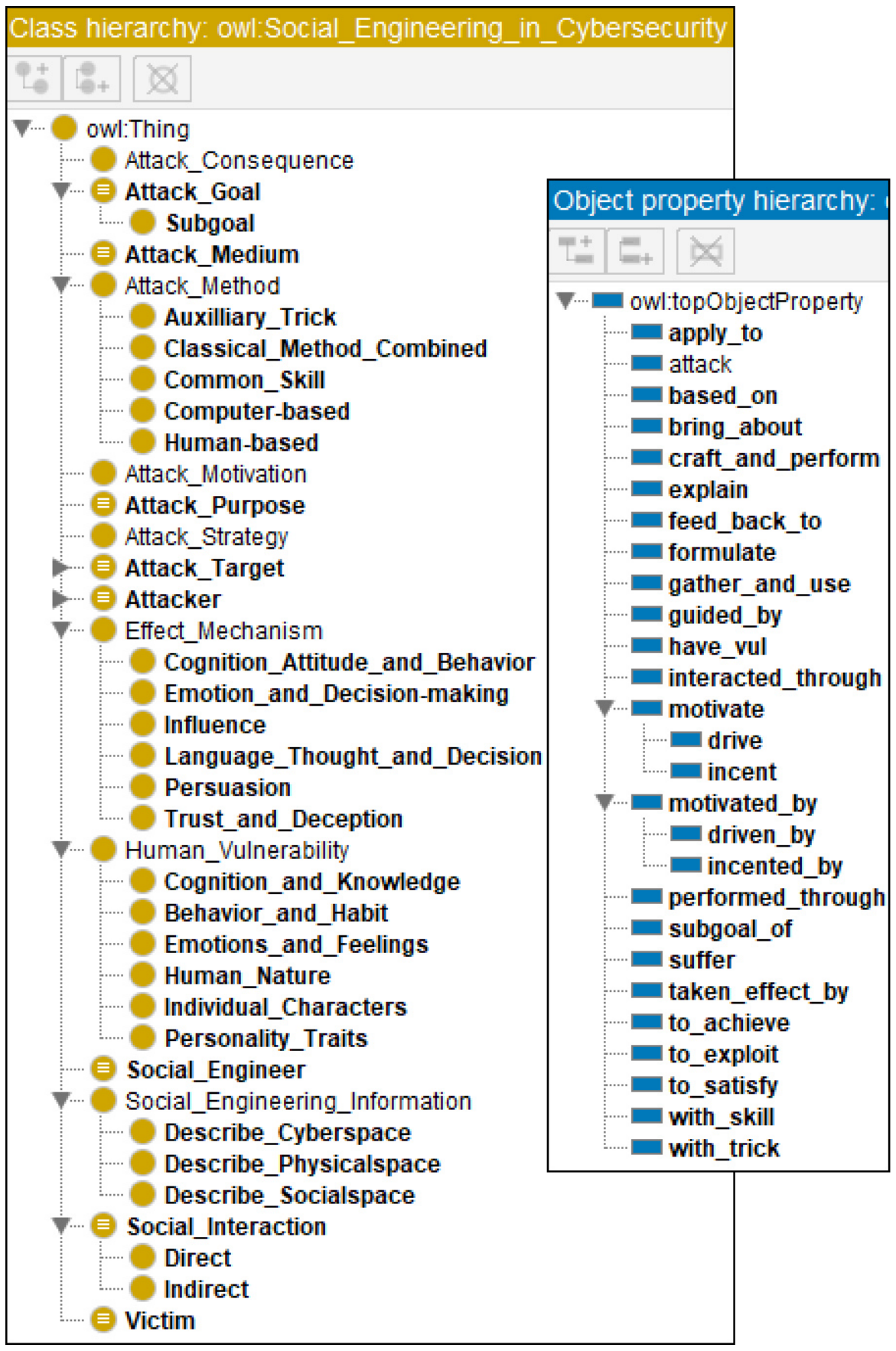

**Figure 14** Overview of concepts and relations defined in Protégé

**Table 2** Other descriptions of concepts (class axioms)

| No. | Concept | Description | Concept |
|---|---|---|---|
| 1 | Attacker | Equivalent To | Social Engineer |
| 2 | Attack Target | Equivalent To | Victim |
| 3 | Attack Goal | Equivalent To | Attack Purpose |
| 4 | Attack Medium | Equivalent To (Entity of) | Social Interaction |

### 3.2 Define relations in the domain ontology

Based on the definitions presented in Section 3.1, we extract 22 kinds of relations among the core concepts. Table 3 shows these relations and their Domain (start), direction and Range (end). Figure 14 (right) shows the implementation of these relations in Protégé, and Table 4 shows the related concepts descriptions set as object property (relation) axioms yet not reflected in the Figure 14 and Table 3.

**Table 3** Define relations among the core concepts

| No. | Concept (Domain) | Relation (→) | Concept (Range) |
|---|---|---|---|
| 1 | Attack Motivation | motivate | Attacker |
| 2 | Attacker | motivated by | Attack Motivation |
| 3 | Attacker | gather and use | Social Engineering Information |
| 4 | Attacker | craft and perform | Attack Method |
| 5 | Attacker | formulate | Attack Strategy |
| 6 | Attack Method | to achieve | Attack Goal |
| 7 | Attack Method | guided by | Attack Strategy |
| 8 | Attack Method | apply to | Attack Target |
| 9 | Attack Method | performed through | Attack Medium |
| 10 | Attack Method | to exploit | Human Vulnerability |
| 11 | Attack Strategy | based on | Social Engineering Information |
| 12 | Attack Target | suffer | Attack Method |
| 13 | Attack Target | have vul | Human Vulnerability |
| 14 | Attack Target | interacted through | Attack Medium |
| 15 | Attack Target | bring out | Attack Consequence |
| 16 | Human Vulnerability | take effected by | Effect Mechanism |
| 17 | Effect Mechanism | explain | Attack Consequence |
| 18 | Attack Consequence | feed back to | Attack Goal |
| 19 | Attack Goal | to satisfy | Attack Motivation |
| 20 | Sub-goal | subgoal of | Goal |
| 21 | Attack Method | with skill | Common Skill |
| 22 | Attack Method | with trick | Auxiliary Trick |

**Table 4** Other descriptions of relations (object property axioms)

| No. | Relation | Description | Relation |
|---|---|---|---|
| 1 | motivate | Inverse Of | motivated by |
| 2 | incent | SubProperty Of | motivate |
| 3 | drive | SubProperty Of | motivate |
| 4 | incented by | SubProperty Of | motivated by |
| 5 | driven by | SubProperty Of | motivated by |
| 6 | incent | Inverse Of | incented by |
| 7 | drive | Inverse Of | driven by |
| 8 | apply to | Inverse Of | suffer |
| | | optional verbose relations | |
| 9 | conduct | Equivalent To | craft and perform |
| 10 | exploited by | Inverse Of | to exploit |

### 3.3 Define other descriptions in the ontology

Besides the axioms descriptions for concepts and relations in Table 2 and Table 4, annotations are optional to facilitate the ontology implementation and many comments (a type of annotation) for instances are added in Section 5.1 to help the instances edition and knowledge analysis.

Here three reasoning rules are defined for simple scenario analysis such as unique attacker, victim and attack consequence. Figure 15. The rule 1 is used to add a new relation: if 1) an attacker crafts and performs certain attack method and 2) the attack method is applied to a target, then a relation "attack" will be created from the attacker to the target (victim). The rules 2 and 3 are used to automatically complete the relations that are not designated explicitly in the instance data but have defined in ontology. This is useful to improve knowledge base and convenient for the instances' creation. The built-in reasoner HermiT can be used to implement the reasoning. For complex attack analysis, these rules might need some adjustments and other reasoning tools can also be used.

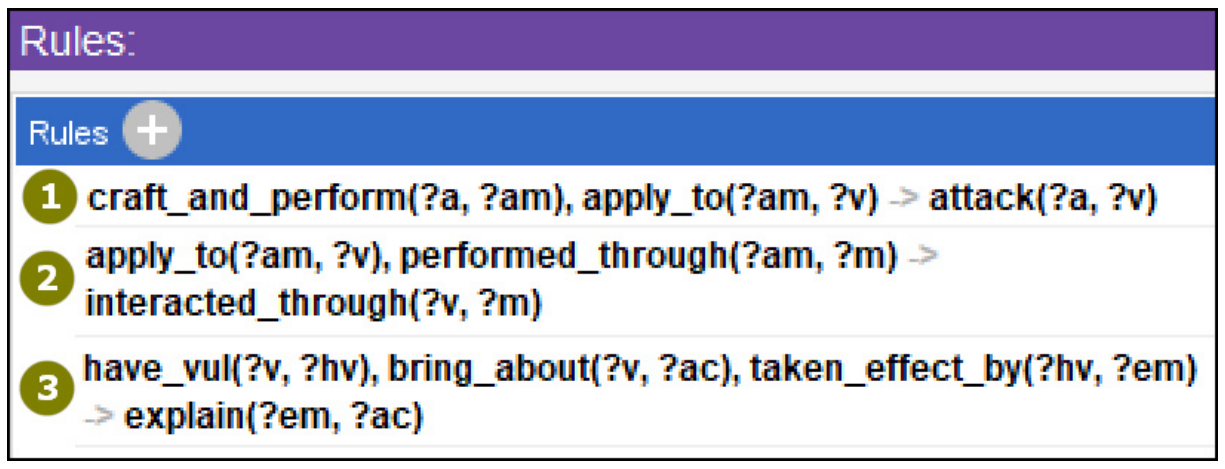

**Figure 15** Rules defined in the ontology

Above is the key material and ontology implementation after the ontology revise and validation. The supplementary material will lead reviewers / independent researcher to reproduce the result.

## 4 Result: domain ontology of social engineering in cybersecurity

Figure 16 shows the domain ontology of social engineering in cybersecurity developed in Protégé [3]. The core concepts and their relations is marked inside the red polygon, the outside shows the taxonomies (also as the labels) used, and the right area is the legend for relations (the directed color connection in the figure). To be intuitive and integrative, Figure 17 presents the ontology in a more clear and concise way.

Overall, 11 core concepts and 22 kinds of relations among them are formally and explicitly encoded / defined in Protégé, together with related description, rules and annotations. For this domain ontology, it can be exported with multiple ontology description language and file formats, such as RDF / XML, OWL /

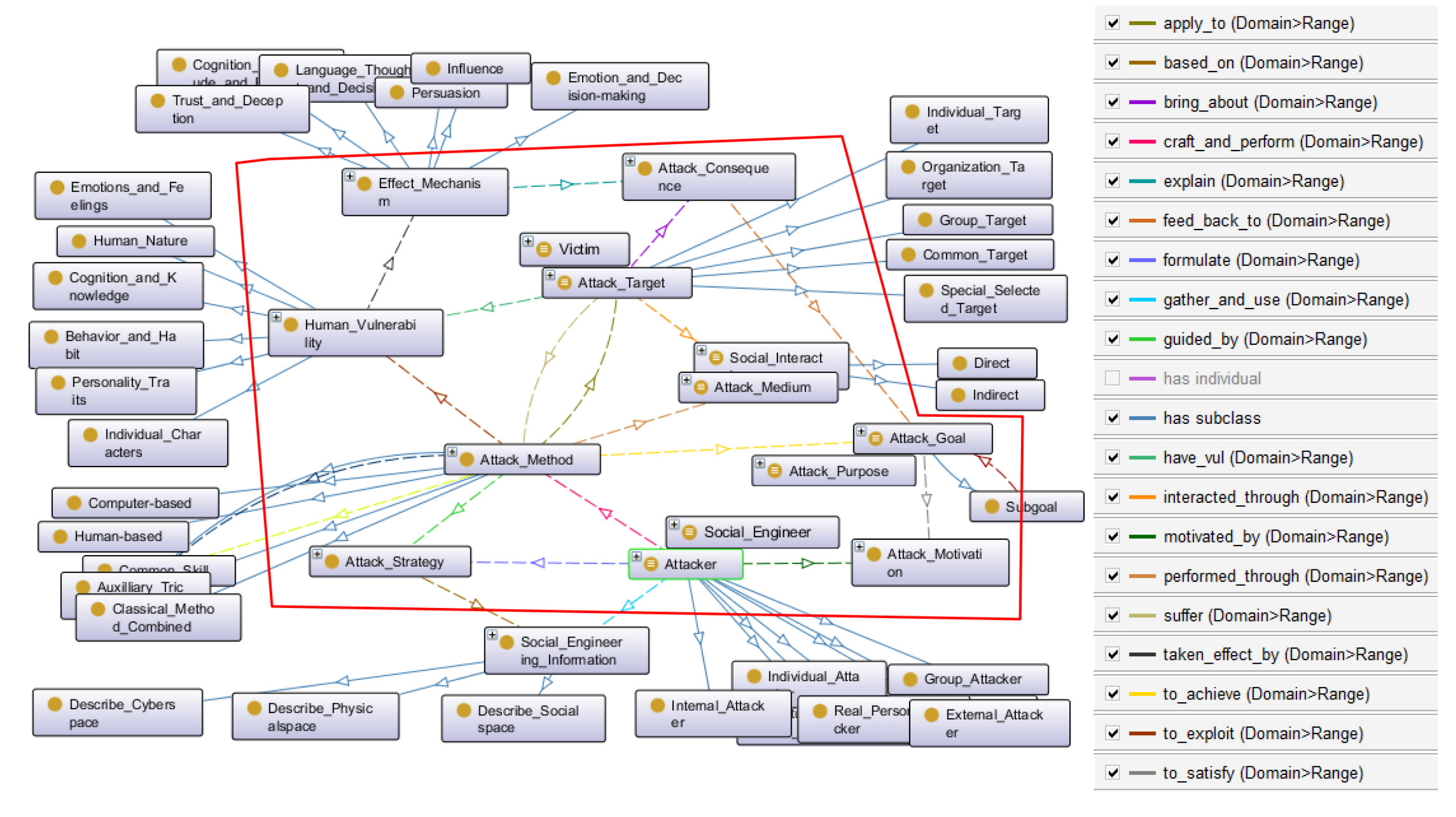

**Figure 16** The domain ontology of social engineering in cybersecurity developed in Protégé

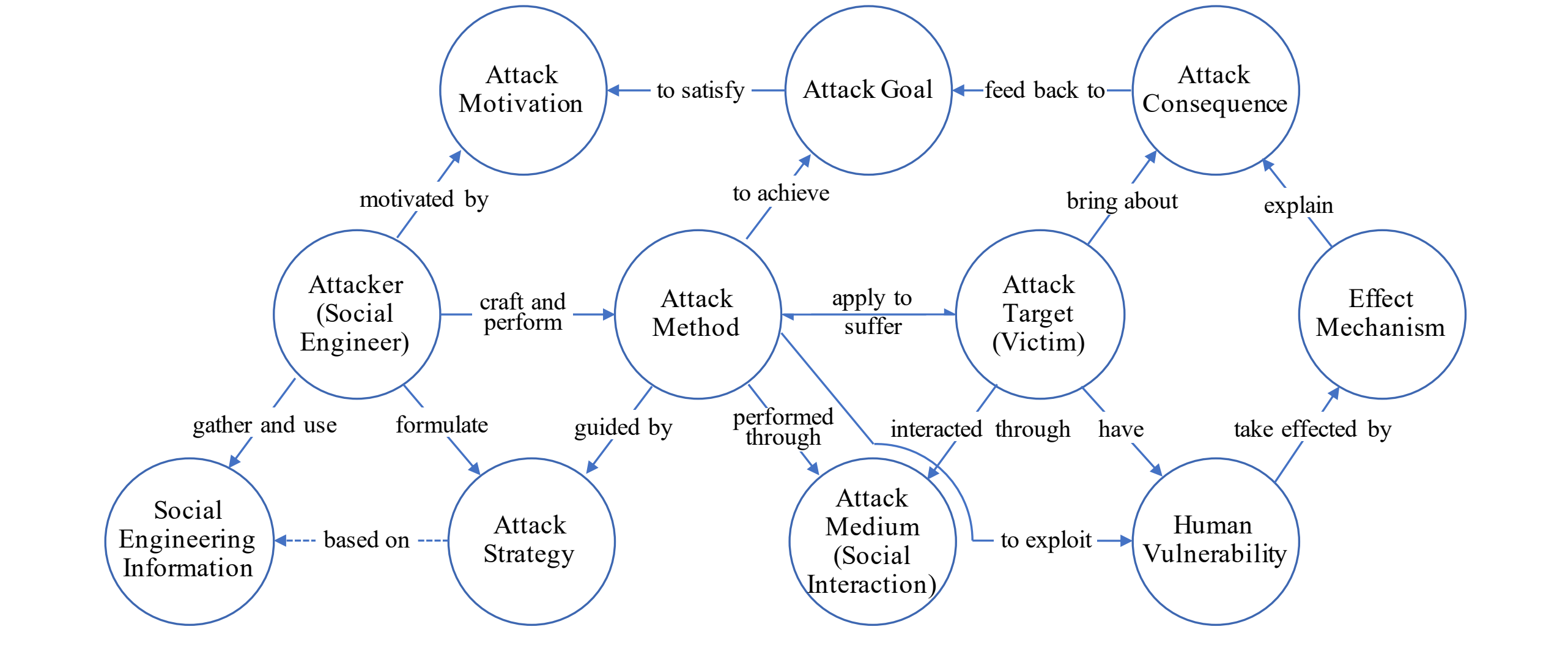

**Figure 17** The domain ontology of social engineering in cybersecurity

XML, Turtle and JSON-LD, to reuse and share the domain knowledge schema.

## 5 Evaluation: knowledge graph application examples

The best way to evaluate the quality of the ontology developed may be problem-solving methods or using it in applications which reflect the design goal [2]. Corresponding to the purpose of the ontology development presented in Section 2, this section evaluates the domain ontology by its knowledge graph application for analyzing social engineering attack scenarios or incidents. First, the ontology serve as a machine processable knowledge schema is used to create the instances, generate the knowledge base and build a knowledge graph. Then, 7 knowledge graph application examples are presented for social engineering attack analysis.

**Table 5** Material of social engineering attack scenarios / incidents adopted from [1, 15] and used to generate knowledge base

| No. | Social Engineering Attack Scenarios / Incidents Description | Human Vulnerabilities | Effect Mechanisms |
|---|---|---|---|
| 1 | **Pretexting**. The attacker attempts to elicit classified or sensitive information from victims (e.g. telephone company operators, motivated by using telephone service without payment) by pretexting via telephone. (1) The attacker makes a prior survey to know better the lingo, organization and victims, and pretexts to be an inner staff (e.g. who is in a trouble) or technical support to elicit information. (2) The attacker requests classified information by pretending to be a cable splicer and pretexting that he is wiring two hundred pair terminals for police. Who would want to refuse a little help to a company man coping with that heavy-duty assignment? She feels sorry for him, she's had bad days on the job herself, and she'll bend the rules a little to help out a fellow employee with a problem. | Credulity or gullibility, sadness, sympathy, the desire to be helpful, agreeableness, kindness and charity, inexperience. | Social responsibility norm and moral duty, (similarity liking and helping), emotions and feelings influence decision-making, ELM, IDT, factors affecting trust. |
| 2 | **Shoulder surfing**. The internal attacker (for security test) pretends to be a maintenance worker to (get access to the target workplace and) contact with the victims. When the victim is not paying attention, the attacker collects information such as username and password by surfing over the victim's shoulder, snooping prominent places such as sticky notes, papers or computers. | Carelessness and thoughtlessness, credulity, gullibility, friendliness, ignorance. | Distraction in persuasion and manipulation, IDT, factor affecting deception and trust, peripheral route to persuasion. |
| 3 | **Vishing and Pretexting**. The attacker (e.g. motivated by financial gain) pretends to be a new employee and convince the targets that he will suffer greatly if the request is not granted. E.g. request the technical support (e.g. Paul) to reset the password of certain account to deal with an urgent task, and further ask a VPN to access from outside. | Guilt, sympathy, the desire to be helpful, friendliness, credulity. | Foot-in-the-door, impression management theory, two routes to persuasion, IDT, cognitive dissonance, ELM, emotions and feelings influence decision-making. |
| 4 | **Vishing and Pretexting**. The attacker (e.g. motivated by financial gain, intellectual challenge) calls a staff of the technical support department to say that the CEO authorized his requesting an urgent VPN channel for a project presentation in another city, and further tells he / she that other staffs did this before, such as Paul. | Fear and dread, conformity, neuroticism, the desire to be helpful, credulity. | Source credibility and obey to authority, diffusion of responsibility, bystander effect, deindividuation in group |
| 5 | **Manipulating conversation**. The attackers (e.g. motivated by fun or pleasure) induce the group conversation to a security topic, one of the attackers discloses his password to discuss whether it is strong enough. If most of the other participants (or attackers) also start disclosing password, the targets are likely to be manipulated to disclose password or other sensitive information. | Conformity, agreeableness, extraversion, credulity, courtesy and humility, diffidence. | Group influence and conformity, social validation, IDT, reciprocity norm, selfdisclosure and rapport relation building, social exchange theory, cognitive dissonance. |
| 6 | **Piggybacking**. An authorized person provides access to an unauthorized person by keeping the secured door open for providing help or other reasons. Most employees do not know every colleague at a (large) organization and will hold a door open for politeness, let alone the attacker is nicely dressed, shoes shined, hair perfect, with polite manner and a smile; victims will less likely to suspect. (e.g. motivated by espionage) | Courtesy, humility, credulity, openness to experience, the desire to be helpful, friendliness, intuitive judgement. | Peripheral route to persuasion, (similarity liking & helping), distraction in persuasion and manipulation, IDT, factors affecting trust, facial expression and deception leakage. |
| 7 | **Trailing and Impersonating**. The attacker (e.g. for security test, personal interest) pretends to be an employee of target organization through suitable disguises such as uniform and printed badge, gaining access to an establishment by following employees who have security card (under the cover of lunch rush at a large corporation). The security guard and employee see in the eye, but he has accustomed to it. In some organizations, the lazy security guards put the access card on the desk for those who forget bringing the access card to pick it up for themselves. | Helpfulness, think set and stereotyping, heuristics thinking and mental shortcuts, intuitive judgement, apathy, indifferent, Ignorance, lazy and sloth. | ELM, peripheral route to persuasion, distraction in persuasion and manipulation, level of need for cognition. |
| 8 | **Baiting**. The attacker (e.g. motivated by competitive advantage) leaves a USB stick containing malicious codes in a location where it is likely to be found by the victims. The outside of the USB stick is the logo of the target organization or attractive icons to lure the victims to pick up and insert into computer. Once inserted, the malicious code may execute automatically. | Curiosity, excitement, greed, conscientiousness, sympathy or the desire to be helpful, inexperience. | (similarity liking and helping), ELM, two routes to persuasion, IDT, emotions and feelings influence decision-making. |
| 9 | **Reverse SE**. The attacker (e.g. motivated by espionage) sends an email using faked address (technical support department) to a new employee informing he / she that "a network test will be conduct recently, and if there is a network failure, please contact xxx". The attacker makes a network fault and waits for the new employee's request. After helping to resolve the problem, the attacker says sincerely "Would you like to do us a favor, just one minute, that completing a survey used for developing a security awareness training program for new employees; nearly 80% of the employees have already done this." "Ok, my pleasure." "Are you aware of our email policies? ... It can be dangerous to open unsolicited attachment ... We need to know your password to evaluate the security awareness of new employees. It is a secure matter" "Okay, it is ..." | Inexperience, intuitive judgement, agreeableness, ignorance, credulity, conformity, the desire to be helpful. | Reciprocity norm, impression management theory, commitment and consistency, framing effect and cognitive bias, language invoke confusion - induce and manipulation, group influence and conformity, diffusion of responsibility, factors affecting trust and deception, IDT. |
| 10 | **Phishing**. The attacker (e.g. motivated by financial gain) sends phishing emails with faked address to inform targets that there is a very low discount coupons of food (or sport event ticket) in a limited time. The email contains tempting food pictures (or passionate sports posters). This lure the targets to click on malicious links (with encoded URL address: att.eg.net), divulge privacy information, etc. | Excitement, happiness, greed, gluttony, surprise, extraversion, impulsion, fear, intuitive judgement. | IDT, peripheral route to persuasion, distraction in persuasion and manipulation, emotions and feelings influence decision-making, scarcity: perceived value and fear arousing. |
| 11 | **Spear Phishing**. The (fired) attacker (e.g. motivated by revenge, financial gain, prank) finds there is some resentment between employees of the target organization through text, images or videos in SNSs, and sends SNSs message or email embedded with malicious code to selected targets, claiming it was a hoax virus that could be forwarded anonymously to someone they didn't like. This may compromise a large group of individuals in the organization. | Disgust, prejudice, anger or wrath, hubris, envy. | Deindividuation, emotions and feelings influence decision-making, neurophysiological mechanism of emotion & decision, micro expression identifying. |
| 12 | **Smishing**. The attacker (e.g. motivated by financial gain, competitive advantage) blocks the target CEO's cell phone signal and sends SMS message to his secretary by faking the CEO's phone number: "I'm in a meeting at another city and couldn't talk on the phone. Encrypt the organization structure table and a contract file to a zip with key *** and send it to xxx@xxx.xxx immediately! Otherwise, we will lose an important business." | Fear and dread, tension, neuroticism, self-love, credulity. | Source credibility and obey to authority, time pressure and thought overloading, emotions and feelings influence decision-making (fear-arousing in persuasion), IDT. |
| 13 | **Trojan attack, honey trap**. The attacker (e.g. motivated by financial gain) puts software in website and implies it is free for downloading and watching porn images or videos. Text marked that "you won't see the seductive images If you don't act." Once the targets opened the link or installed the software, the attacker's computer or mobile device is compromised. | Lust, greed, excitement, curiosity, impulsion, intuitive judgement. | IDT, emotions and feelings influence decision-making, peripheral route to persuasion, distraction in persuasion and manipulation, indirectness of thinking and negative expression in language. |
| 14 | **Water-holing**. The attacker (e.g. motivated by financial gain) finds that the targets usually, regularly, will or are likely to visit certain websites, and then infects these websites with malicious code waiting for the targets' trigger. The targets will be compromised e.g. when visit the websites, download software (malware) or click (malicious) links. | Fixed-action patterns, behavioral habits of site-visiting, think set and stereotyping. | IDT, factors affecting trust and deception, social and organizational trust theory. |
| 15 | **Whaling attack**. A spear phishing attack directed specifically at high-value targets such as senior executives, CEO or CFO. The attacker (e.g. motivated by financial gain) craft the whaling baits such as emails and websites are highly customized and personalized, in which the target's name, job title, job responsibility, internal phone numbers, organizational logos, email footer and other relevant information are incorporated. And the attack is usually context-aware, e.g. "... the xxx business meeting/ conference in your schedule needs you to register and confirm the registration using the attached software (trojan horse or back door with encoded domain address: att.eg.net)". | Heuristics and mental shortcuts, intuitive judgement, carelessness, thinking set or stereotyping, credulity. | ELM, the central route to persuasion, the peripheral route to persuasion, time pressure and thought overloading, factors affecting deception and trust, integrative model of organizational trust. |

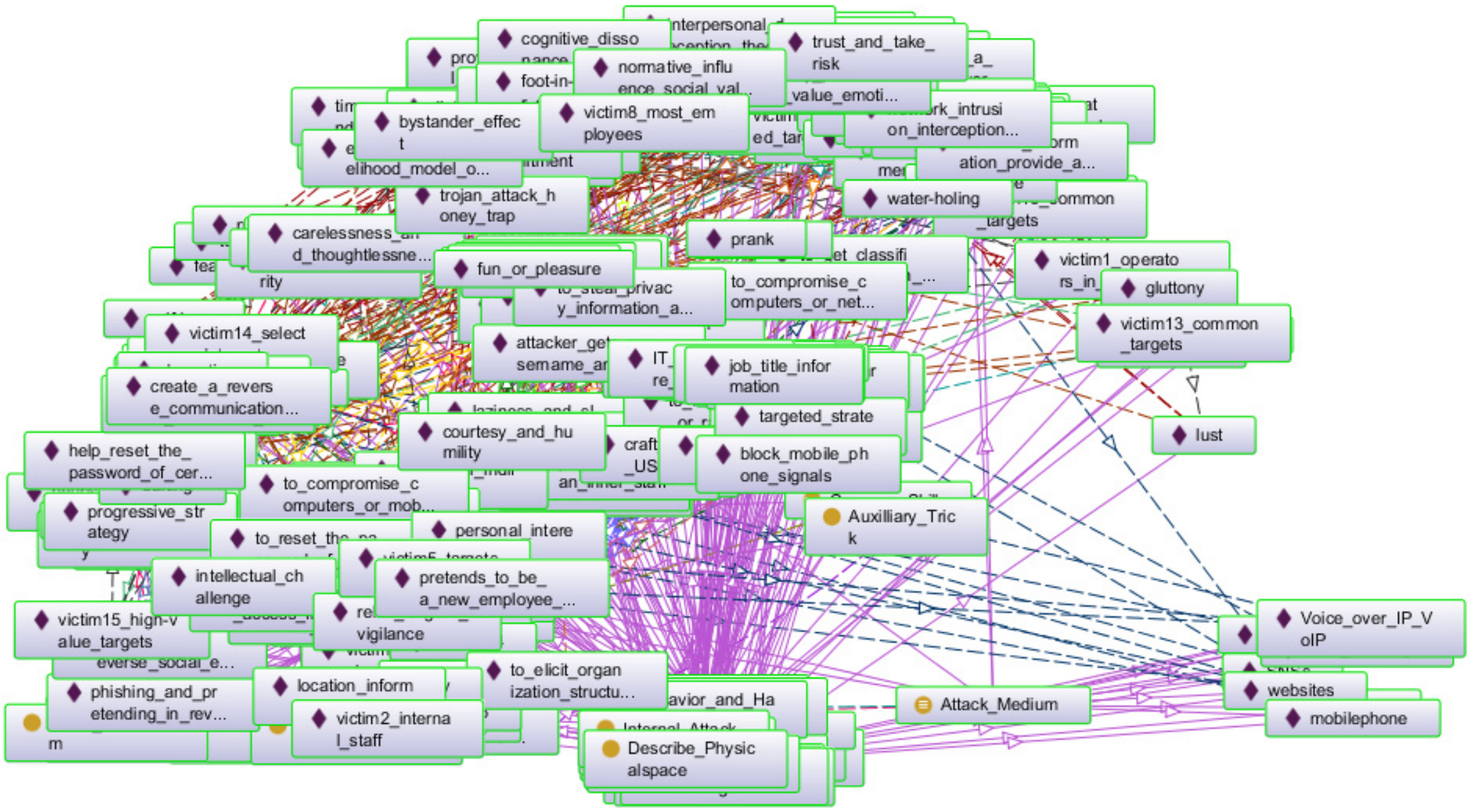

**Figure 18** The overview of the knowledge base generated in Protégé

## 5.1 Create instances, knowledge base and knowledge graph

An ontology together with a set of instances organized by the knowledge schema defined by the ontology constitutes a knowledge base, which further serve as the data source of a knowledge graph. For this paper, a dataset of social engineering attack scenarios that contains the necessary instance classes such as attacker, victim / target, human vulnerability, social interaction (medium) and attack goal is in demand. Yet there is not such a public dataset available now. Thus, the attack incidents and typical attack scenarios described in work [1] and [15] are adopted and expanded as material to create instances for each concept defined in the ontology and build the knowledge base. Overall, 15 attack scenarios (Table 5) in 14 social engineering attack types are used to generate a relatively medium-small size knowledge base.

The instances and their interrelations described in every attack scenario are dissected and edited in Protégé also, since it is convenient to check the data consistency and revise errors according to the ontology. In this process, we add many comments (for instances of attacker, attack method and victim) to assist the instances creation and knowledge analysis. Figure 18 shows the overview of the knowledge base in Protégé. A total of 224 instances are created in the knowledge base. [4]

[4] The implementation file was submitted as supplementary material (SEiCS-Ontology+instances-inferred.owl) for review.

Due to the limited functionality of Protégé for data analysis and visualization, we select Neo4j (community-3.5.19) [16] as the tool to display the knowledge graph and analyze social engineering attacks. Neo4j is easier and faster to represent, retrieve and navigate connected data. And the Neo4j CQL (cypher query language) commands are declarative pattern-matching, which is in human-readable format and easy to learn.

There are mainly two steps to migrate data from Protégé to Neo4j. First, export the ontology and instances in Protégé to RDF/XML or OWL/XML file, with the reasoner enabled to infer and complete the knowledge according to the axioms and rules defined. Then, import the RDF/XML [4] file into Neo4j by the plugin neosemantics (version 3.5.0.4). The detailed scripts and commands used to build the knowledge graph is submitted as supplementary material.

According to the statistic in Neo4j, 1785 triples were imported and parsed, and 344 resource nodes and 939 relations were created in the whole knowledge graph. Figure 19 shows the knowledge graph consist of all instances nodes and their interrelations. The legend for node color is in the left bottom.

In the knowledge graph, the relations *craft and perform, apply to, to exploit, have vul, bring about* among nodes *attacker, attack method, victim, human vulnerability, attack consequence* are colored with red, to abstract and denote an attack occurrence (Figure 19), for the convenience of attack analysis.

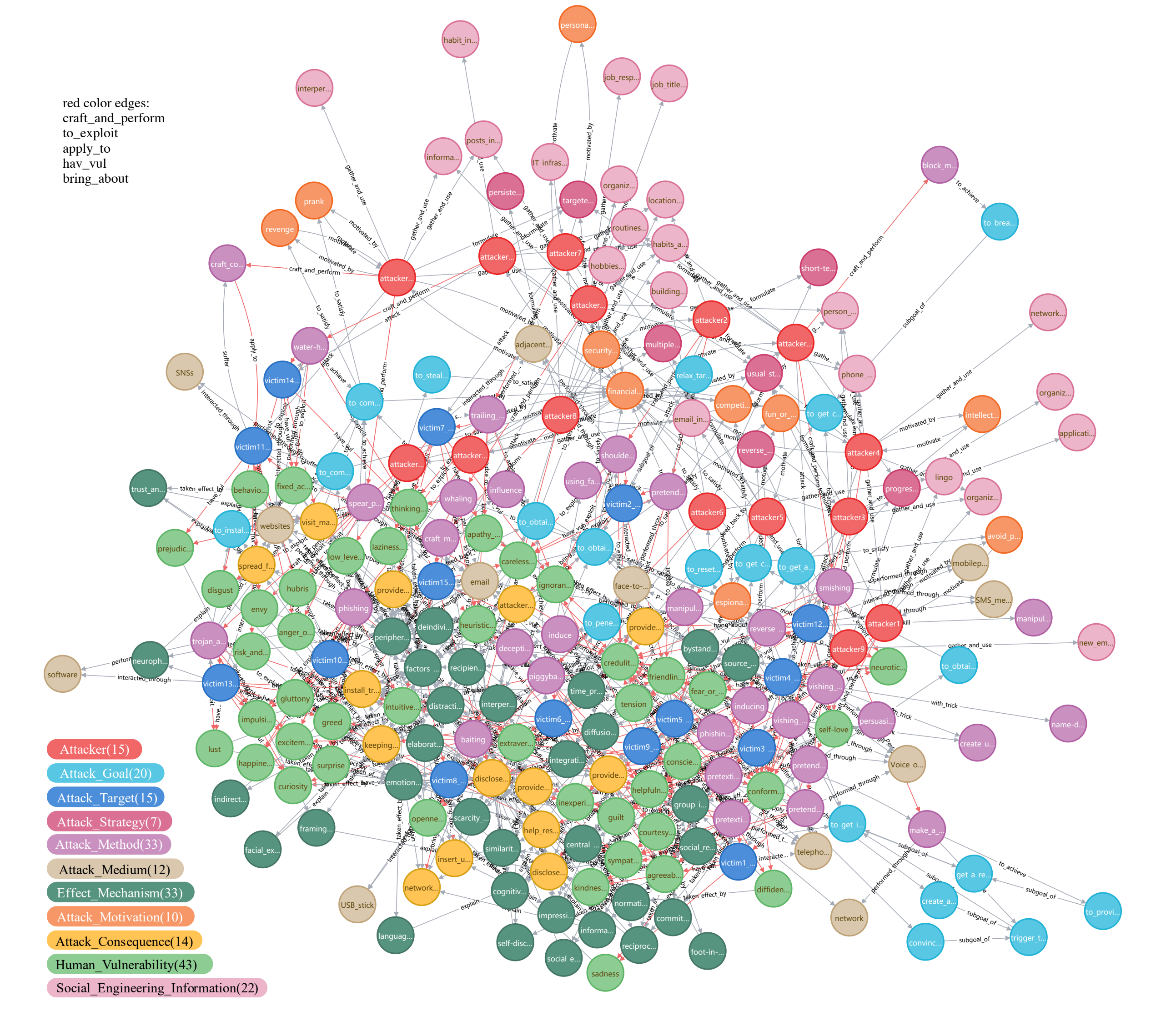

**Figure 19** The knowledge graph generated in Neo4j

## 5.2 7 knowledge graph application examples

By virtue of the domain ontology and knowledge graph, there are at least 7 application examples (in 6 patterns) available to analyze social engineering attack scenarios or incidents. [5]

### *5.2.1 Analyze single social engineering attack scenario or incident*

The components of a specific social engineering attack scenario can be dissected into 11 classes of nodes with different color. These nodes are interconnected and constitute an intuitive and vivid knowledge graph. By this way, the security researchers can get an insight of an attack quickly from the whole to the part.

[5] All the CQL scripts for these application were submitted as supplementary material for review.

A case in point is the knowledge graph of attack scenario 9 (a reverse social engineering attack) as Figure 20 shows. The left part (of area 2) depicts the contents surrounding the attacker: the attacker9 motivated by espionage to gather and use information about organization structure, new employee and email address; formulate reverse and progressive strategy; craft and perform (red arrow) multiple attack methods to elicit password or other sensitive information, or get access or help to breach cybersecurity. Goal and subgoals in area 1 form an attack tree structure, which enables to describe the multi-step attacks in progressive strategy or other complex attack scenarios. The middle part (area 2) depicts the attack mediums through which the attack methods are performed, and also the interaction form with targets (victims). The right part depicts the nodes related to victim: the victim9 brings about certain attack consequences, due to he / she has vulnerabilities such as conformity, inexperience and helpfulness, which (are exploited by attack methods and) are taken effect by mechanisms displayed in the right edge nodes. Some relations (suffer, to exploit, explain) are not displayed here to get a clear view, which can be returned by adding CQL expressions, clicking the node (expand / collapse relations) or using the setting "connect result nodes".

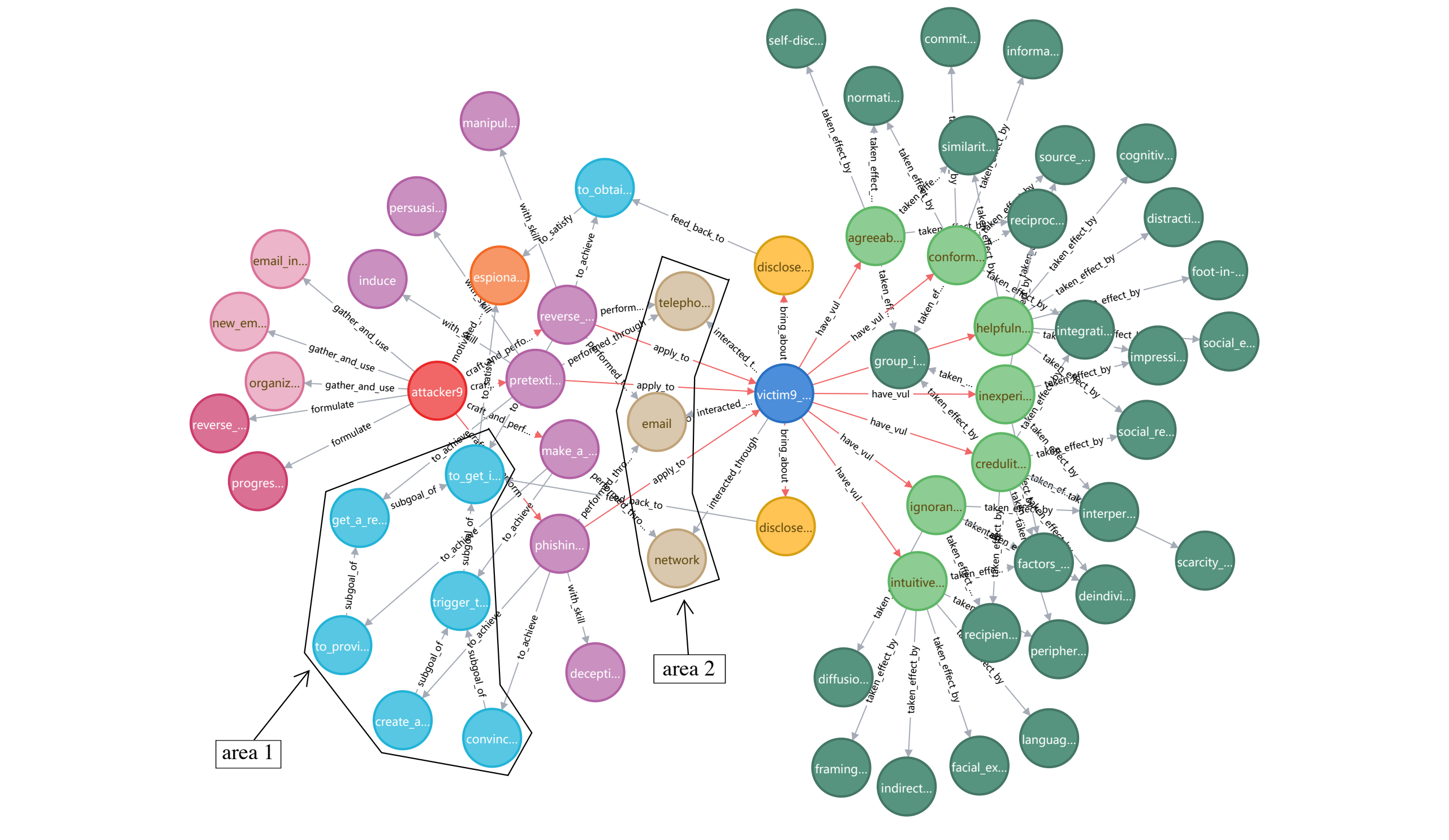

**Figure 20** Analyze single social engineering attack scenario (e.g. scenario 9) by knowledge graph

### 5.2.2 *Analyze the most exploited human vulnerabilities*

As one of the confrontational focuses between social engineering attack and defense, human vulnerability is what attackers want to exploit and what defenders / victims want to eliminate or mitigate. Knowing the frequently exploited human vulnerabilities is of great significance for social engineering defense. The exploited frequency for each human vulnerability in the knowledge base can be counted and ranked by CQL expressions (MATCH, COUNT, ORDER). Figure 21 extracts the top 3 human vulnerabilities most exploited by various kinds of attack methods: credulity, helpfulness and conformity. This suggests that these human vulnerabilities should be watched out in security-related issues and paid more attention in defense measures such as security awareness training.

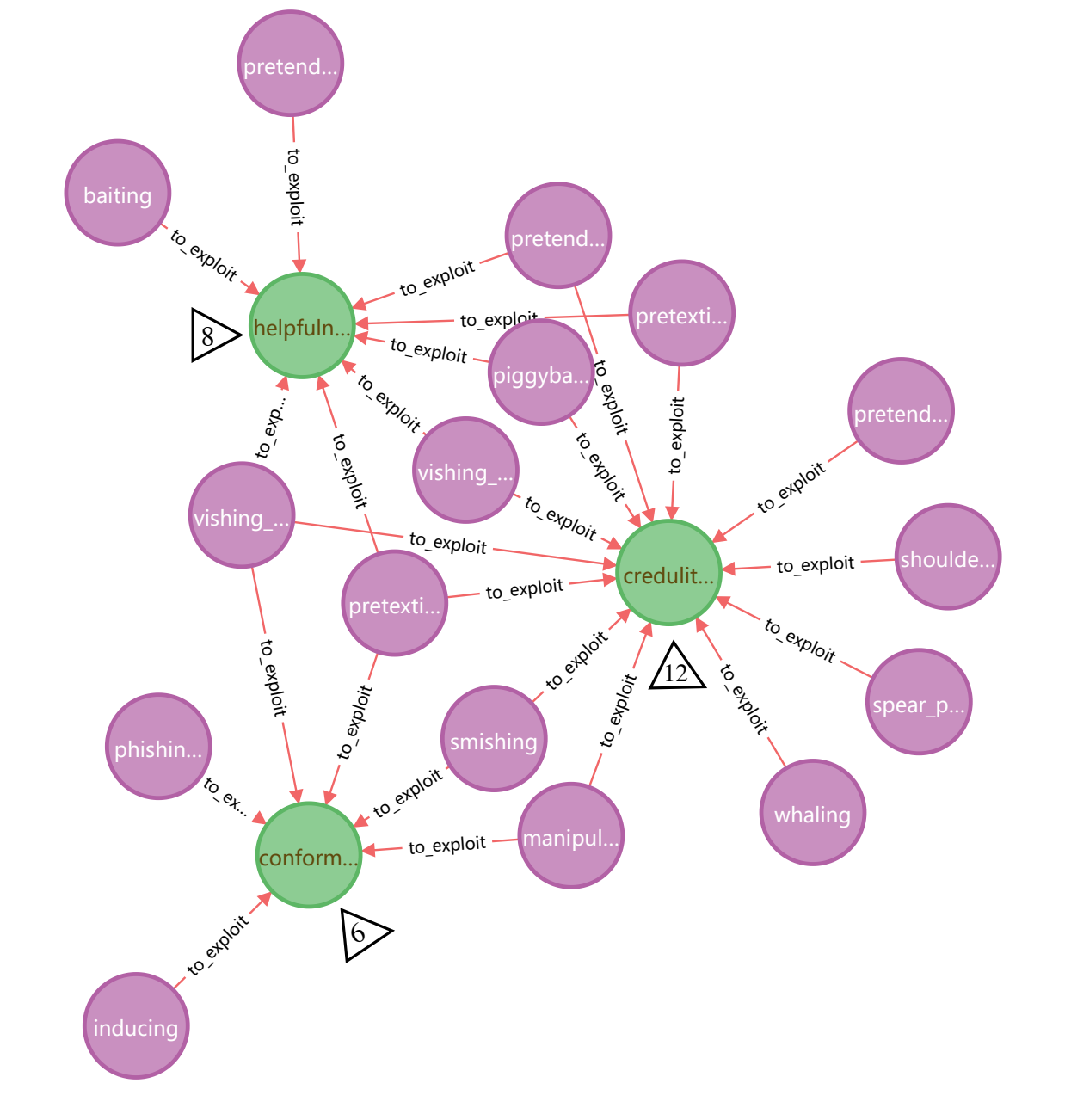

**Figure 21** Find the most exploited (top 3) human vulnerabilities

#### *5.2.3 Analyze the most used attack mediums and interaction forms*

Similar to the analysis pattern in Section 5.2.2, the statistic analysis of attack mediums and interaction forms can be executed to get an understanding of where the social engineering attacks are frequently occurred. Figure 22 presents the top 3 mediums most used to perform social engineering attack in the knowledge base: email, website and telephone. This reflects that many social engineering attacks are performed through network and electronic communication, meanwhile reminds us to beware social engineering threat when using these communication mediums.

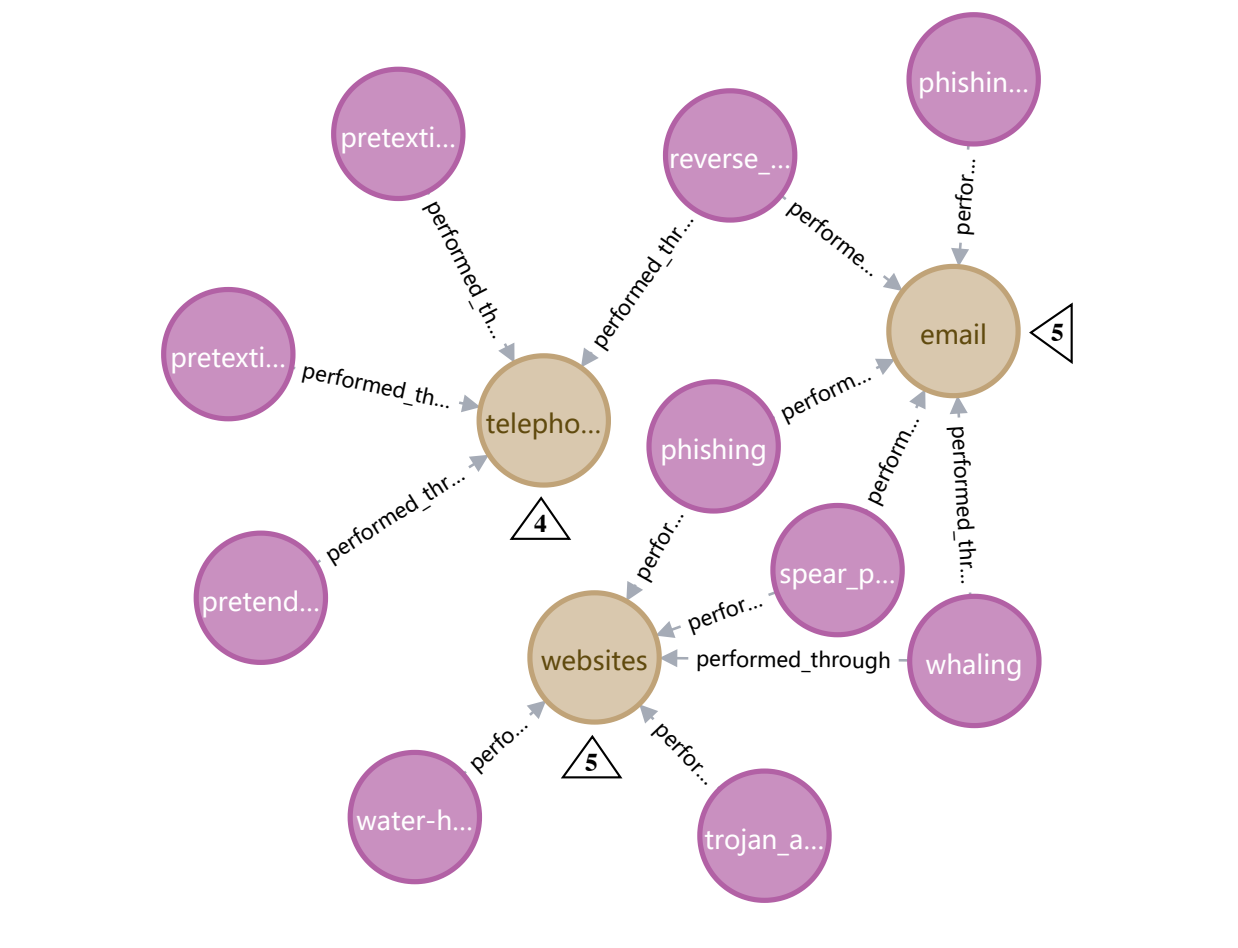

**Figure 22** Find the most used (top 3) attack mediums and social interaction

#### *5.2.4 Find additional (potential) threats for victims (targets)*

For specific victim (target), knowledge graph can be used to find additional (potential) threats beyond the given scenario. The following analysis pattern can be extracted from the domain ontology and attack scenario analysis:

1. **if** $(v1) \rightarrow (hv)$ in S1 **and**
   $(a2) \rightarrow (am2) \rightarrow (hv) \leftarrow (v2)$ in S2
2. **then**
   $(a2) \rightarrow (am2) \rightarrow (hv) \leftarrow (v1)$ is feasible

Namely: the attacker *a2* can also employ the attack methods *am2* to attack victim *v1* (i.e. exploited the victim *v1's* vulnerabilities *hv*), if a victim *v1* has certain human vulnerabilities *hv* and exploited in scenario *S1* meanwhile the *hv* are found also exploited in another scenario *S2* by attacker *a2* through attack method *am2*.

Figure 23 shows this application where victim7 serves as an example. It depicts that the victim7 has five human vulnerabilities and exploited by attacker7 in scenario7; besides, three of these vulnerabilities can be also exploited by another 5 pairs of attacker and attack method. In short, for victim7 there are 5 additional and potential attack threats, and precautions should be taken against them.

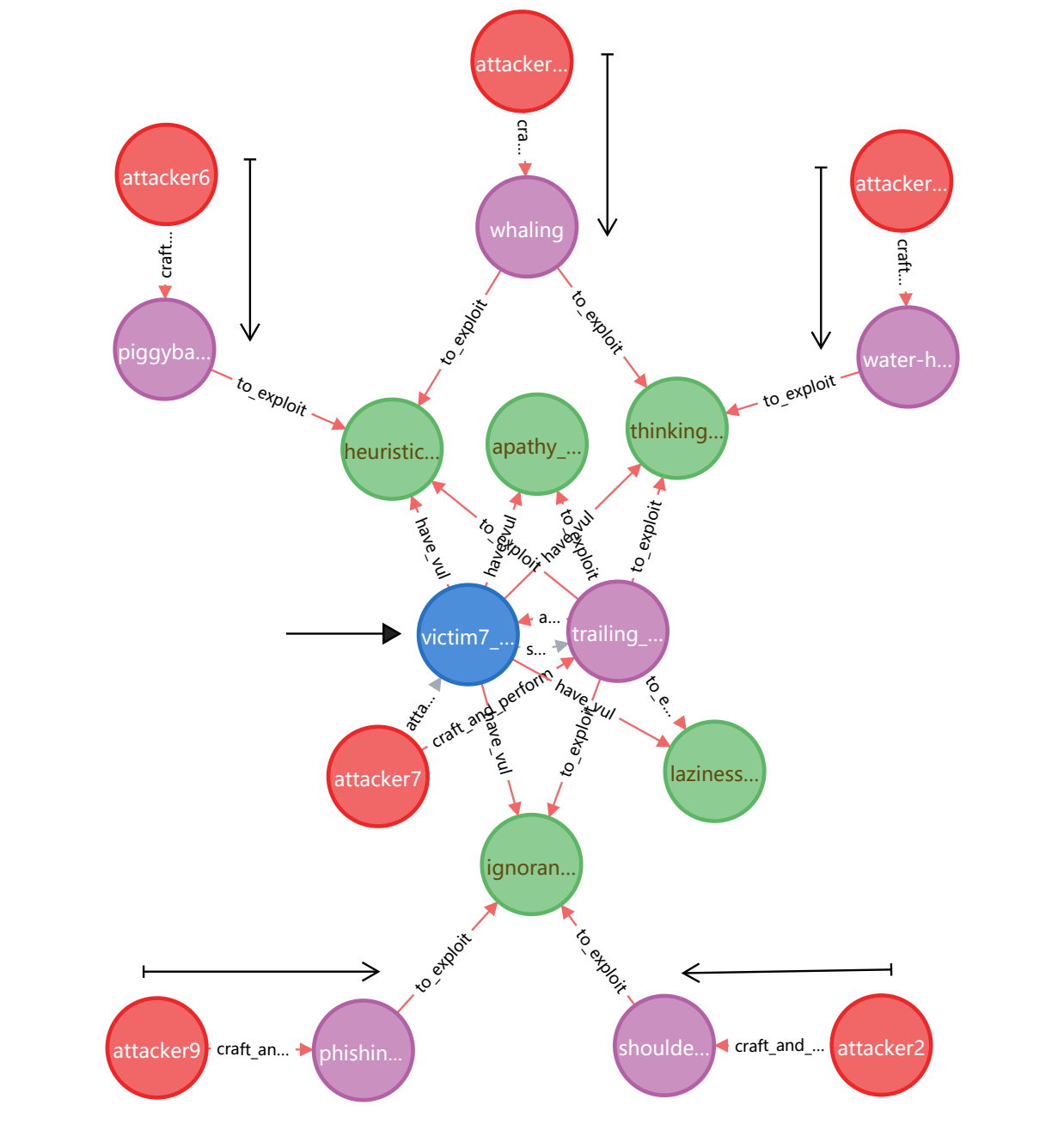

**Figure 23** For specific victim, find additional threats beyond the given scenario

To evaluate this and the latter two analysis patterns, we extracted all the undirected and acyclic graphs (among red color edges) from an attacker to a victim in the knowledge graph. This treatment generated a clear labeled dataset, meanwhile avoided the subjectivity in the process of labeling. In total, 345 reachable paths (i.e. attack paths) were labeled.

Among these attack paths, 177 (attacker, attack method) pairs are labeled. For all the 15 victims, this analysis pattern find 156 new (attacker, attack method) threat pairs beyond the 21 pairs described in Table 5. Besides, the above analysis pattern recalls 176 pairs without wrong cases. The recall rate is 99.43% and the F1 score is 99.71%. One pair was omitted due to one attack method's edges *to exploit hv* were divided and assigned to other attack methods in the same scenario.

#### *5.2.5 Find potential targets for attackers*

For specific attacker, knowledge graph can be used to find additional or potential targets beyond the given scenario. Similar to the previous analysis pattern, the following logic was extracted:

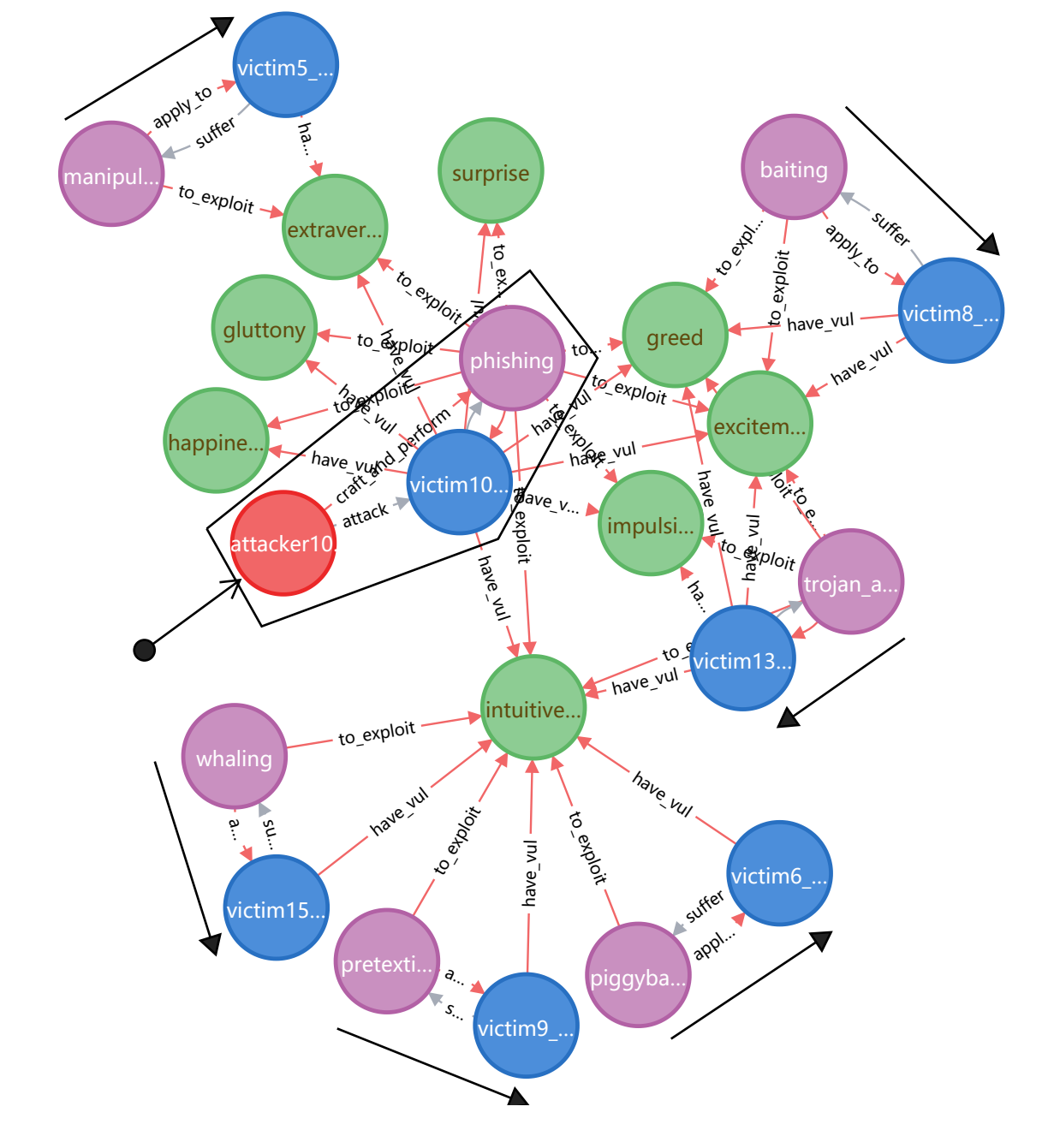

**Figure 24** For specific attacker, find potential targets (victims) beyond the given scenario

1. **if** *(a1)* → *(am1)* → *(hv)* ← *(v1)* in S1 **and** *(am2)* → *(hv)* ← *(v2)* in S2
2. **then** *(a1)* → *(am1 or am2)* → *(hv)* ← *(v1)* is feasible

Namely: the victim *v2* can be also attacked by the attacker *a1* through attack method *am1 or am2*, if a victim *v1* has certain human vulnerabilities *hv* and exploited by attack method *am1* crafted by attacker *a1* in scenario *S1* meanwhile the victim *v2* is found also has the same vulnerabilities *hv* in scenario *S2* exploited by attack method *am2*.

Figure 24 shows this application where attacker10 serves as an example. It presents that the attacker10 crafts and performs phishing to exploit victim10's vulnerabilities in scenario10; moreover, another 6 targets have the same vulnerabilities that victim10 has and can be also exploited by attacker10 through phishing (or attack methods in other scenarios). In brief, 6 potential targets are found for attacker10. For practice, it is helpful to notify all the potential targets if attacker10 or phishing is a serious security threat. If this is a penetration testing, Figure 24 will offer testers more attack targets and attack methods.

For all the 15 attackers, this analysis pattern find 123 new exploitable targets beyond the 15 victims described in Table 5, and 156 new (attack method, targets) pairs beyond the 21 pairs described in Table 5. This analysis pattern recalls 176 (attack method, targets) pairs without wrong cases. The recall rate is 99.43% and the F1 score is 99.71%. One pair was omitted due to one attack method's edges *to exploit hv* were divided and assigned to other attack methods in the same scenario.

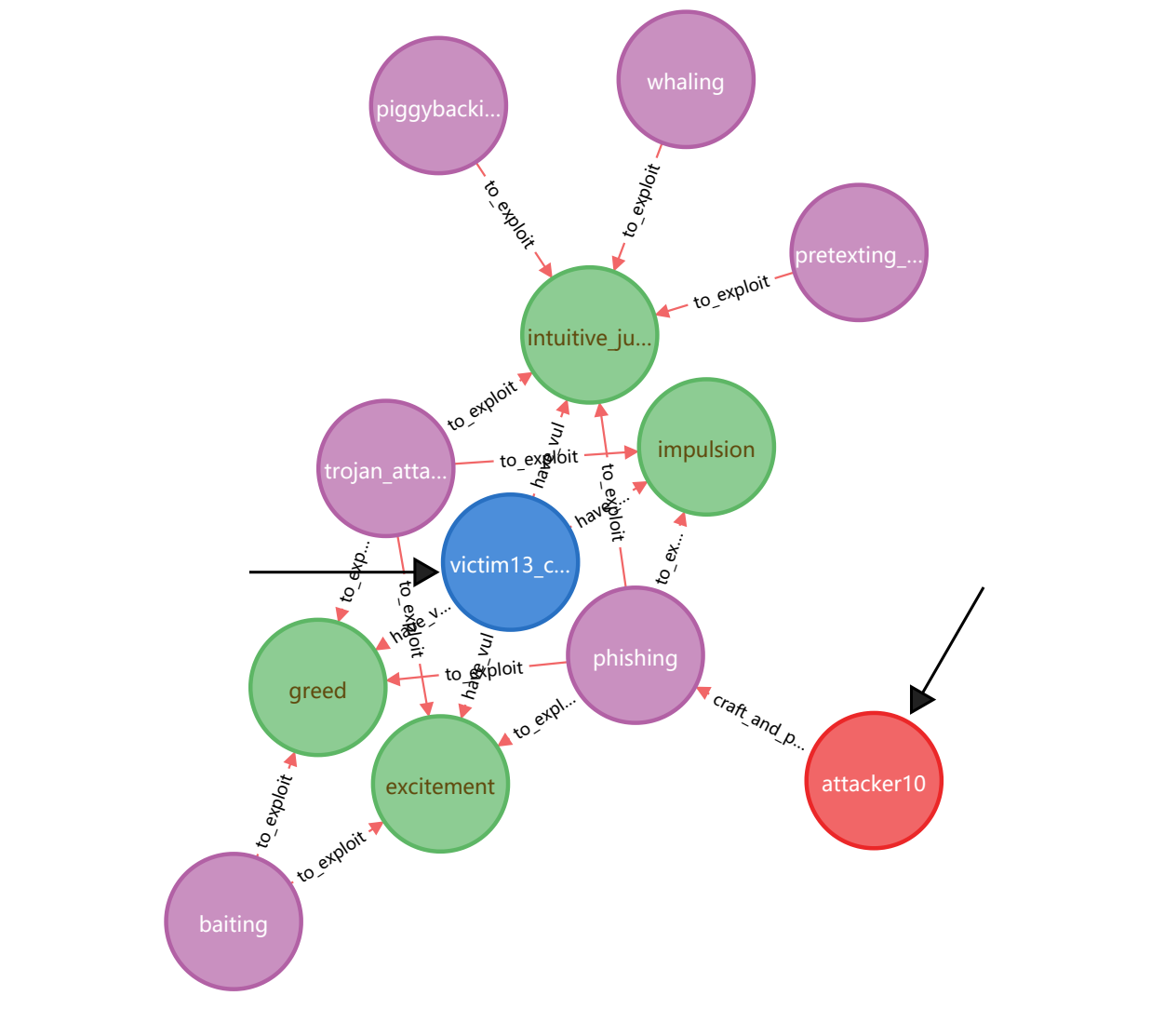

**Figure 25** For specific attacker and victim, find potential attack paths and methods

### *5.2.6 Find paths from specific attacker to specific target*

For specific attacker and specific victim which are not in the same attack scenario, knowledge graph can be used to check or find feasible attack paths and potential attack methods. This is a combination of the previous two analysis patterns, and the following pattern was extracted:

1. **if** *(a1)* → *(am1)* → *(hv)* in S1 **and** *(v2)* → *(hv)* in S2
2. **then** *(a1)* → *(am1)* → *(hv)* ← *(v2)* is feasible

Namely, the attack path from attacker *a1* to target *v2* is feasible, if attacker *a1* can successfully exploit human vulnerability *hv* by attack method *am1*, meanwhile the target *v2* is found has the vulnerability *hv*.

Figure 25 shows this application where attacker10 and victim13 serve as the examples. The following 4 attack paths is extracted from the knowledge base: *(attacker10)-[craft and perform]* → *(phishing)-[to exploit]* → *(4 human vulnerabilities)* ← *[has]-(victim13)*. In addition, another 5 attack methods that exploit the victim13's vulnerabilities but not within the attack paths are also presented in Figure 25. These methods are potentially available for attacker10 to reach victim13.

For all the 15 attackers and 15 targets, this analysis pattern find 251 new attack paths beyond the 94 paths

| | (A, AM) | (AM, T/V) | AP |
|---|---|---|---|
| Labeled | 177 | 177 | 345 |
| Recall | 176 | 176 | 344 |
| Recall rate | 99.43% | 99.43% | 99.71% |
| F1 score | 99.71% | 99.71% | 99.85% |

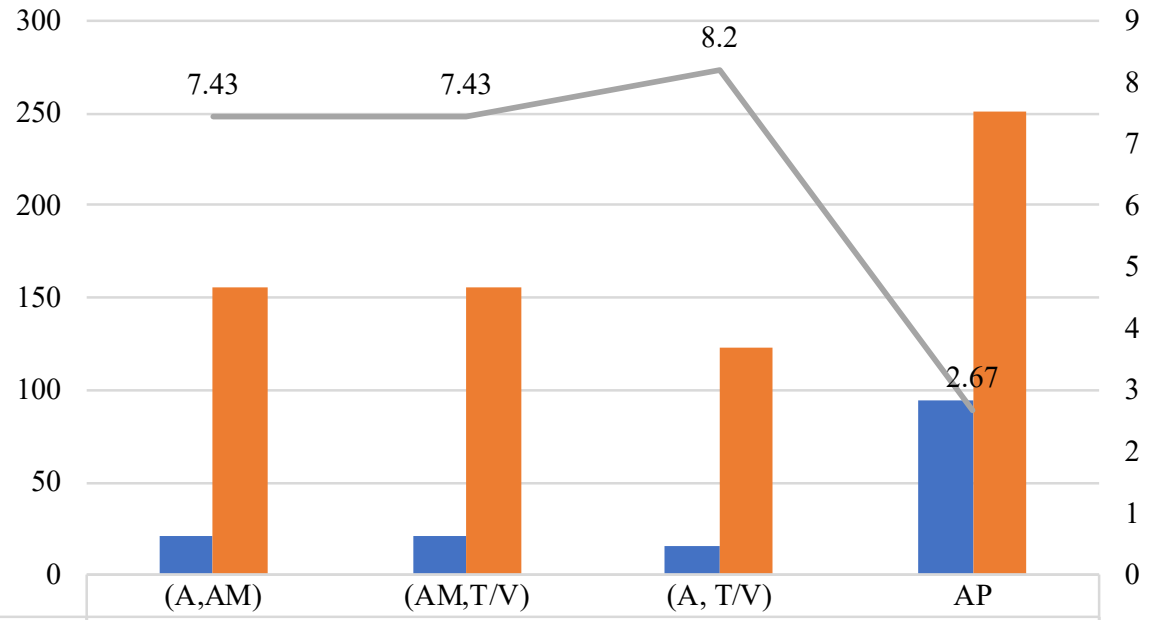

| | (A,AM) | (AM,T/V) | (A, T/V) | AP |
|---|---|---|---|---|
| existed in scenarios description | 21 | 21 | 15 | 94 |
| new find beyond the scenarios description | 156 | 156 | 123 | 251 |
| folds | 7.43 | 7.43 | 8.2 | 2.67 |

A: attacker
AM: attack method
T/V: attack target / victim
AP: attack paths

**Figure 26** Experiment results and statistic analysis of Section 5.2.4, 5.2.5 and 5.2.6

described in Table 5, and 123 new (attacker, targets) pairs beyond the 15 pairs described in Table 5. For all 345 labeled attack paths, this analysis pattern recalls 344 attack paths without wrong cases. The recall rate is 99.71% and the F1 score is 99.85%. One attack path was omitted due to one attack method's edges *to exploit hv* were divided and assigned to other attack methods in the same scenario.

Figure 26 summarizes the experiment results and statistic analysis of Section 5.2.4, 5.2.5 and 5.2.6.

### *5.2.7 Analyze the same origin attack*

In general, the attack method *am1 and am2* are similar or related if they have some common features; *am1 and am2* might be launched by the same attacker if they have certain crucial common features, e.g they point to the same domain address controlled (by attacker). Further, *am1 and am2* is likely to be same-origin and the attacker *a1 and a2* is likely in the same attack organization, if above *(am1, am2)* are launched respectively by two different attackers *(a1, a2)* who are motivated by the same motivation *m* to attack different victims *(v1, v2)* who have the same affiliation. Based on above cognition or assumption, Figure 27 shows the knowledge graph application example to analyze same origin attack.

Besides returning the graph existed in the knowledge base, new relations and nodes can be created. A new relation "same affiliation" is created between victim10 and victim15, since they both have the data property "affiliation" with the equal value. There is a potential relation "same origin attack" between whaling and phishing nodes, because in the whaling attack Trojan horse or back door with encoded domain address "att.eg.net" is used meanwhile this address is also found in the malicious link of phishing attack. Furthermore, due to attacker15 and attacker10 have the same motivation "financial gain" and victim15 and victim10 in the same "Company A", given all these, it can be inferred that these two scenarios compose a same-origin and organized attack. Thus, we create new relation "same origin attack" between the two attack method nodes and relation "in the same organization" between the two attacker nodes.

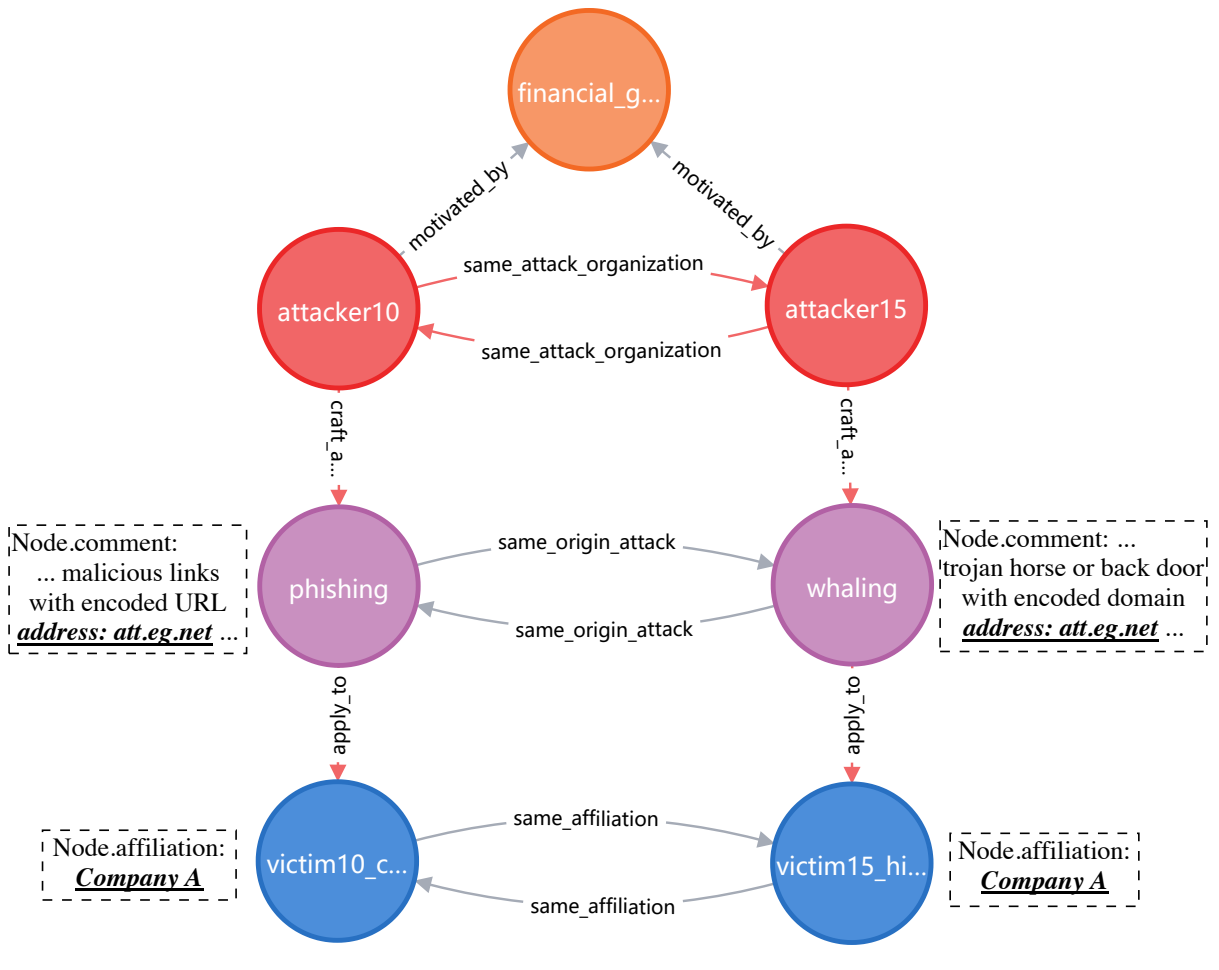

**Figure 27** Analyze the same origin attack by knowledge graph

## 6 Discussion

There are some studies related to social engineering ontology. Simmonds et al. [17] proposed a conceptualization / ontology for network security attacks, in which components (access, actor, attack, threat, motive, information, outcome, impact, intangible, system administrator) are included. Although some components (e.g. actor, motive, information) are similar to concepts in this paper, the ontology [17] focuses on network security (and access control), which cannot be used to describe social engineering domain. Oosterloo [18] presented an ontological chart, in which concepts such as attacker, threat, risk, stakeholder and asset are involved. But this chart is served as a model to summarize and organize aspects related to social engineering risk management, and the purpose is not a formal and explicit description of concepts and relations in social engineering domain. Vedeshin [19] discussed three phases (orchestration, exploitation, and compromise) of social engineering attacks, in which some classes (such as target, actor, goal, techniques, medium, execution steps and maintaining access) are discussed. However, this taxonomy is used to classify different social engineering attacks. Mouton et al. [20] described an ontological model of social engineering attack consisted of six entities: social engineer, target, medium, goal, compliance principles and techniques. However, the concept definitions of these entities were not presented and the relations among these entities were also not specified. That is, it does not constitute a domain ontology. Besides, the social engineering definition in [20] is proposed form the perspective of persuasion, which describes only a part of social engineering [1]. As another result, the model does not include some important entities (e.g. human vulnerability) and aspects (e.g. deception and trust). Tchakounté et al. [21] discussed a certain spear phishing scenario / flow and its description logic (DL), yet other social engineering attack types were not involved. Li and Ni [22] discussed the difficulty to distinguish social engineering attacks (methods) collected from six studies. They identified some core concepts to characterize social engineering attack by aligning these concepts with existing security concepts, and then provided a description logic for a security ontology and attack classification. In the security ontology, *social engineer, social engineering attack, human and human vulnerability* were respectively aligned as subclass of *attacker, attack, asset and vulnerability*; another two concepts *attack media and social engineering techniques* were also included. However, *human* is the *target* yet not the *asset* that social engineering attacks aim to *harm*, and according to their text and ontology implementation, social engineering *attack* and *technique* seem to refer the same concept. This might be reasons why the concepts' relations in their work were not aligned. Besides, the domain ontology of social engineering is not the focus of study [22], and the above six (or five) concepts are not sufficient to analyze relatively complex social engineering attack incidents / scenarios. Alshanfari et al. [23] gathered some terms related to social engineering and attempted to organize them by Protégé using method described in [2]. However, the terms were extracted only from 30 publications from 2015 to 2018 and only three entity classes (attack type, threat and countermeasures) were presented, in which some terms are just related to the class yet are not the instances of it (e.g. *guilt, websites* in *attack type*; *sensitive information, password* in *threat*). Besides, relations among these classes were not described clearly. Thus, this work is mainly oriented to the terms and classification. Nevertheless, we would like to appreciate above works and other researchers who make efforts in this field.

We develop a domain ontology of social engineering in cybersecurity and conducts ontology evaluation by knowledge graph application.

- The domain ontology describes what entities significantly constitute or affect social engineering and how they relate to each other, provides a formal and explicit knowledge schema, and can be used to understand, analyze, reuse and share domain knowledge of social engineering.
- The 7 analysis examples by knowledge graph not only show the ontology evaluation and application, but also present new means to analyze social engineering attack and threat.
- In addition, the way that *1) use Protégé to develop ontology, create instances and knoledge base 2) and then employ Neo4j to import RDF/OWL data, optimize knoledge base and construct knoledge graph for better data analysis and visualization* also provides a reference for related research.
- In the ontology, some taxonomies (subclasses) or relations might be verbose or omitted. But as mentioned before, subclass name will be converted to node labels and inverse relations can facilitate the knowledge retrieval, and therefore, users can add or delete them based on specific application requirements.
- The material of attack scenarios and the data of ontology+instances offer a dataset can be used for future related research. The knowledge graph dataset (224 instances nodes, 344 resource nodes and 939 relations of 15 attack scenarios) seems small. Yet it covers 14 kinds of social engineering types, and the 6 kinds of analysis patterns have demonstrated the various feasibilities of the proposed ontology and knowledge graph in analyzing social engineering attack and threat.
- To the best of our knowledge, this is the first work which completes a domain ontology for social engineering in cybersecurity, and further provides its knowledge graph application for attack analysis.

Due to the complexity of social engineering domain, the ontology seems impossible perfect in the only once establishment. We throw out a brick to attract a jade and look forward superior studies by researchers in this field.

## 7 Conclusion

This paper develops a domain ontology of social engineering in cybersecurity, in which 11 concepts of core entities that significantly constitute or affect the social engineering domain together with 22 kinds of relations among these concepts are defined. It provides a formal and explicit knowledge schema to understand, analyze, reuse and share domain knowledge of social engineering. Based on this domain ontology, this paper builds a knowledge graph using 15 social engineering attack incidents / typical scenarios. The 7 knowledge graph application examples (in 6 kinds of analysis patterns) demonstrate that the ontology together with the knowledge graph can be used to analyze social engineering attack scenarios or incidents, to find (the top ranked) threat elements (e.g. the most exploited human vulnerabilities, attack mediums), to find potential attackers, targets and attack paths, and to analyze the same origin attacks.

**Author details**

[1]School of Cyber Security, University of Chinese Academy of Sciences, Beijing, CN. [2]Beijing Key Laboratory of IoT Information Security Technology, Institute of Information Engineering, Chinese Academy of Sciences, Beijing, CN.